\documentclass[]{aa}
\usepackage[varg]{txfonts}
\usepackage{graphicx}
\usepackage[hidelinks]{hyperref}
\usepackage{natbib}
\usepackage{amssymb}
\usepackage{siunitx}
\usepackage[version=4]{mhchem}
\bibpunct{(}{)}{;}{a}{}{,} % to follow the A&A style

\begin{document}

\title{Spatial-temporal forecasting the sunspot diagram}

%\subtitle{Spatial-temporal forecasting of the sunspot butterfly diagram using local state space reconstruction}

\author{Eurico Covas
\inst{1}}

\institute{Independent researcher, 38 Somerfield Road, London N4 2JL, United Kingdom,
\email{eurico.covas@mail.com}}

\date{Received 16 June 2016 / Accepted 28 March 2017}

\abstract{}
{We attempt to forecast the Sun's sunspot butterfly diagram in both space (i.e.\ in latitude) and time, instead of the usual
one-dimensional time series forecasts prevalent in the scientific literature.}
{We use a
prediction method based on the non-linear embedding of data series in high dimensions. We use this method
to forecast both in latitude (space) and in time, using a full spatial-temporal series of the
sunspot diagram from 1874 to 2015.}
{The analysis of the results shows that it is indeed possible to reconstruct the overall  shape and amplitude of the spatial-temporal
pattern of sunspots, but that the method in its current form does not have real predictive power.
%It also seems to show
%that existing methods that just look at forecasting the sunspot's maximum count for the next cycle
%or attempt to predict the shape of the next cycle could
%be misleading. This is because our results show that different forecasts with distinct likeness to the observed future butterfly diagram
% (as measured by the so called structural similarity measure)
%can have an indistinguishable forecasted maximum and/or overall shape to the actual one.
We also apply a metric called structural similarity to compare the forecasted and the observed 
butterfly cycles, showing
that this metric can be a useful addition to the usual root mean square error metric when analysing the efficiency of different prediction methods.

}
{We conclude that it is in principle possible to reconstruct the full sunspot butterfly diagram for at least one cycle using this approach and that this method and others
should be explored since just looking at metrics such as sunspot count number or sunspot total area coverage is too reductive 
 given the  spatial-temporal dynamical complexity
of the sunspot butterfly diagram. However,  more data and/or an improved approach is probably necessary to have true predictive power.}


\maketitle

\section{Introduction}
\label{introduction}

Sunspots are both spatial and temporal physical phenomena visible at the surface of our Sun (and also other stars), that appear darker against
the very bright background. The Sun's surface has an average temperature of around
5780K, while by contrast sunspots have a temperature between 3000K and 4500K. This is why sunspots
appear as dark spots compared to the remaining solar surface.
Sunspots are created by strong concentrations of magnetic
fields, these inhibit convective movements at the solar surface and this reduction in convection then reduces the
temperature \citep[see][and references therein]{2004A&G....45d..14P}. Sunspots usually show up on the surface  in pairs, each one having
an opposite magnetic polarity \citep{1908ApJ....28..315H,1999ApJ...525C..60H}.

Records going back to 800 B.C.\ show that astronomers in ancient China were observing and
recording sunspots \citep{1980AmJPh..48..258S,1989QJRAS..30...59M}. Much later in England, an English monk named John of Worcester
made the first sketch of sunspots on 8 December 1128 \citep{john1995chronicle}.
After the invention of the telescope in the early 1600s, observations of sunspots become
more common and regular. In 1843 a German astronomer, Samuel Heinrich Schwabe, revealed the rise and fall of the yearly sunspot
count;  this marks the discovery of the sunspot cycle \citep{1844AN.....21..233S}. At first, it was
thought that the sunspot cycle was strictly periodic, and Samuel Heinrich Schwabe in 1843 estimated the period to around 10 years \citep{1844AN.....21..233S}.
Later it was found that the sunspot cycle is not periodic, or even quasi-periodic, but follows what seems to be a low
dimensional chaotic oscillation \citep{1981ComAp...9...85R, 1988ssgv.conf...69W, 1990RSPTA.330..617W, 1991JGR....96.1705M, 2006A&A...449..379L, 2009SSRv..144...25S,2014SSRv..186..525A}, i.e.\ one
which never repeats itself exactly.
Subsequently, in 1848, Rudolph Wolf introduced the relative
sunspot number, $R$, which now takes his name, confirmed Schwabe's discovery, and through a study of daily observations
of sunspots found that the average length or average period of a solar cycle was about
11 yrs.\ \citep{1852MNRAS..13...29W,1859MNRAS..19...85W}. Other measures were introduced later, such as the
`smoothed monthly mean sunspot number' \citep{1961says.book.....W, 1987JGR....9210101W, 1994ssac.book.....W}.
The analysis of the length of the sunspot cycle showed that this chaotic oscillation has a 
recurring time between 10 and 12 or so years.  Interestingly, these oscillations in the mean cycle
period have been associated with possible changes in the global climate \citep{Wilson2006, 2008PASA...25...85W}.

Sunspot area measurements since 1874 describe the
total surface area of the Sun covered by sunspots
at a given time. These measurements contain both time and latitude information \citep{1980SoPh...68..303Y, butterfly}.
It is these sunspot area measurements that this article uses and analyses. Both the sunspot area coverage and the several
sunspot measures show a similar low dimensional chaotic behaviour with an average period or recurring time of 11 years.
These two measures, sunspot area and sunspot count, have been shown to be closely related \citep{2006STIN...0620186W}, in fact to
be piecewise linearly related. The area data is considered to have more physical significance because it is the
sunspot area that is related to the total magnetic flux  at the solar surface \citep{2007SoPh..240...17P,2014SSRv..186..105E, 2014ApJ...792...12M}.

As well as the cycle itself, it was found that sunspots seems to emerge at the mid-latitudes ($\pm 35$ degrees),
but as the sunspot cycle reaches a maximum (in both number of sunspots and sunspot area measures)
the sunspots move to lower latitudes \citep{1859MNRAS..19...85W}. Near the minimum of the cycle, sunspots appear even closer to the equator,
and as a new cycle starts, sunspots again start emerging at the mid-latitudes. This pattern is called the
`butterfly' diagram and was first discovered by Edward Ma\"under and Annie Ma\"under in 1904 \citep{1904MNRAS..64..747M}
\citep[see also][and references therein]{1971JBAA...81..270G}, There is also a small overlap of the two cycles, where sunspots for the
new cycle emerge while the previous cycle terminates.

In addition to the low dimensional chaotic behaviour and the butterfly diagram, the sunspot cycle series
have other very interesting dynamical features. The Ma\"under Minimum is one of them; it
is the name used for the period  approximately between 1645 to 1715 when sunspots became extremely rare.
The term sunspot minimum was first introduced by John Eddy in a 1976 article \citep{1976Sci...192.1189E}.
Other astronomers before Eddy had also named the period after Annie and Edward Ma\"under. This absence of sunspots
was not an error due to the lack of observations or to the quality of the telescopes at the time as there was evidence for a cycle before the minimum \citep{1978A&A....66...93W}.
Sp\"orer noted that during one 28-year period within the Ma\"under Minimum (1672--1699),
observations recorded fewer than 50 sunspots, much fewer than  is typically observed today.
A possible explanation for the coexistence of these two modes of behaviour, a weakly chaotic quasi-cycle and a switching
to minima, could be that the Sun's magnetic field is affected by intermittency \citep{1993A&A...274..497C, 1997PhRvE..55.6641C,
1999ASPC..178..173T, 1997jena.confE..55P, 1999PhRvE..60.5435C, 2001Chaos..11..404C, 2001SoPh..199..385C, 2003IJBC....8.2327O, 2004ApJ...616L.183C, 2005SoPh..229..345C,
2009SSRv..144...25S, 2016AJ....151....2D} or by a two-mode switching \citep{2016MNRAS.456.2654W}.

This and other earlier  sunspot cycle minima are also clearly visible in the records of cosmogenic isotopes, e.g.\ \textsuperscript{14}C
and \textsuperscript{10}Be \citep{1998SoPh..181..237B}, which are proxy or tracers of sunspot and/or solar activity.
There is also evidence for the Ma\"under Minimum and possibly other minima in tree ring analysis \citep{1980Natur.286..868S, 1980Sci...207...11S,
2012PNAS..109.5967S}.
Interestingly, the Ma\"under Minimum coincided with a period of lower than average European temperatures,
hinting at a possible Sun--Earth climate connection \citep{1978Natur.276..802T, 1991Sci...254..698F, 2006JASTP..68.2053D,9780309265645}.
In addition to this possible relationship,
another connection was noticed by Edward Sabine, who noted that the average period of the sunspot cycle
was similar in value to the period of changes of Earth's geomagnetic activity,
giving birth to the study of solar-terrestrial connections which we now call `space weather' \citep{1851RSPT..141..123S, 1852RSPT..142..103S,
1979P&SS...27.1001S,1983SoPh...89..195E}.
In fact,  solar activity can affect human activity in multiple ways. Solar storms can affect spacecraft electronics \citep{
2000AdSpR..26...27W,2011SpWea...9.6001C,2013EGUGA..1510865W}, increase the radiation exposure for humans in space \citep{2003A&AT...22..861B, 2006GMS...165..367T, 2009SunGe...4...55C,2011AtmEn..45.3806S}, affect and shut down electric grids \citep{2005SpWea...3.8C01K}, and produce subtle variations in Earth's
climate \citep{1991Sci...254..698F,1995JATP...57..835L}, among others.
The impact of the sunspot cycle on the climate and on geomagnetic activity, together with its direct impact on
human activity makes the
forecasting of the solar magnetic activity and/or the sunspot cycle of great importance.

An entire industry
has been created around forecasting the sunspot cycle, using multiple methods, too many to mention in this article
(for several
reviews, see e.g.  \citealt[][and references therein]{1999JGR...10422375H},
and also   \citealt{1995SoPh..159..371K, 2003SoPh..218..319U,2012SoPh..281..507P}). These predictions or forecasting methods can be
divided between pure mathematical methods (see e.g. \citealt{1999SoPh..189..217K, 2005ARep...49..495O, 2005SoPh..231..167O}),
which ignore the physics underlying the time series, and methods based
on some physical underlying mechanism \citep[see e.g.][and references therein]{2005GeoRL..3221106S, 2005GeoRL..32.1104S, 2006GeoRL..33.5102D},
using mechanisms that can explain the solar cycle such as dynamo theory (see e.g. \citealt{9780198512905}, and
for reviews see \citealt{2003ASPC..286...97O, 2003A&ARv..11..287O}). There are also techniques based on a combination of these
two methods (see e.g. \citealt{2006GeoRL..3318101H, 2003SoPh..213..203D}).

A lot of emphasis and hard work
\citep{1987GeoRL..14..632S, 1996GeoRL..23..605S, 1991SoPh..132....1L,
2005GeoRL..32.1104S, 2006GeoRL..33.5102D, 2007MNRAS.381.1527J, 2009SSRv..144..401H, 2014SoPh..289.1815P}
has gone into forecasting the sunspot number cycle, and in particular the sunspot number at the
solar cycle maximum. For the current solar cycle, which is the 24{th} solar cycle since counting started in 1755
and which began in late 2008
there has been an extensive body of literature trying to forecast the sunspot number at the maximum. 
The sunspot number at the maximum is a reductive metric,
 and  in fact so many articles are published every year that no matter what number is calculated as the maximum, 
 there will always be an article that  matches the observed number within a reasonable error interval.  
 This is not because of the lack of scientific
value of these works, but because the maximum sunspot number over a cycle is a simplified metric, being zero-dimensional. Even the sunspot number shape across the whole of the cycle
can be said to be a simplified metric as well, being one-dimensional. To make the point further, we note that the smoothed sunspot
number\footnote{For a description of the currently used version (version V2.0 ) of the smoothed sunspot number see \url{http://solarscience.msfc.nasa.gov/predict.shtml}.} reached a peak of 116.4 in April 2014. However,
we can see from the results of the American Geophysical Union (AGU) meeting in December 2008
that there was a space filling set of forecasts, as can be seen in the `piano plot'
presented by W.\ D.\ Pesnell who has surveyed the scientific literature for forecasts of the cycle 24 maximum sunspot number
\citep[see][]{pianoplot,2012SoPh..281..507P}. This plot illustrates that  the sunspot number (or other metrics such as the average sunspot number or total sunspot area coverage)
are metrics which do not seem good enough to answer the question of `What is the best forecasting method for sunspot activity?' In other words,
the sunspot number maximum (and the sunspot number time-series and its relatives) are functions
of the total four-dimensional (three space and one time dimensional) physical data set, which are too low dimensional
to be able to evaluate and compare the accuracy of different forecasting methods. Basically, important information gets lost and we cannot decide between
forecasting methods.

\begin{figure*}[htp]
\centering
\includegraphics[width=17cm]{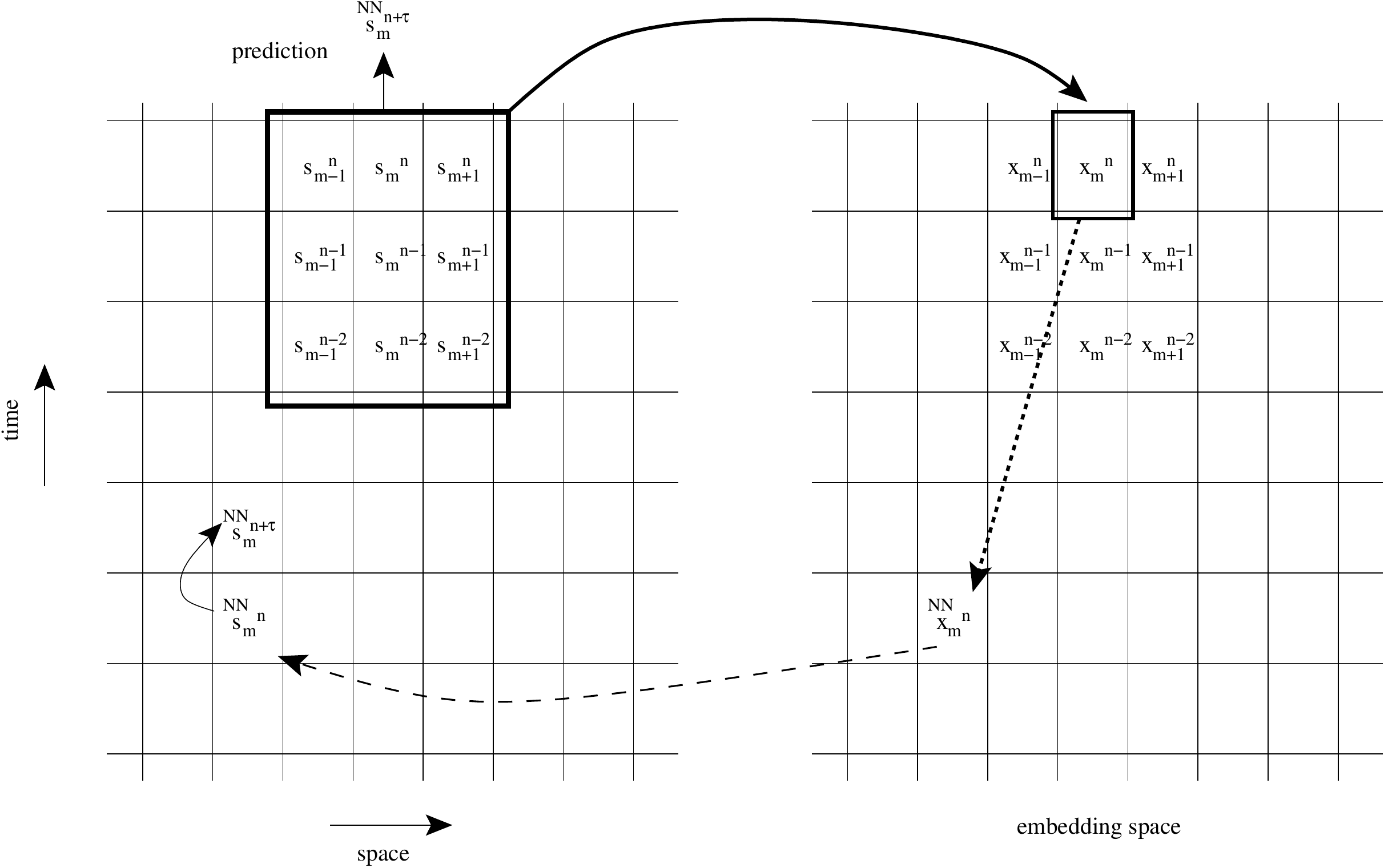} \caption{Representation of forecasting method. An embedding space is constructed using
space and time delays, then  the nearest Euclidean neighbour is found. Once found, it can be used as an approximation to deduce the next time prediction in the real
space-time original space.}
\label{forecasting_xfig}
\end{figure*}

In this article we propose what we believe is  a more general approach, by attempting to forecast the full sunspot butterfly diagram, i.e. 
performing a two-dimensional
forecast (time versus latitude), based on the full data set for the sunspot areas and its latitudinal position, from 1874 to 2015\footnote{We use the
data set publicly available in \url{http://solarscience.msfc.nasa.gov/greenwch/bflydata.txt } \citep{butterfly}.}.
 We use a method based on
local state space reconstruction that was applied previously to spatial-temporal financial data and published with Filipe C.\ Mena \citep{Covas}.
 This is, however, not the first time the forecast or reconstruction of the full sunspot butterfly space-time diagram has been attempted.
\citet{2011A&A...528A..82J} used the full sunspot diagram data series to uncover statistical relationships or correlations between the
latitude, longitude, area, and tilt angle of sunspot groups against the cycle strength and cycle phase. Once those relationships
were established, they were then able to reconstruct the sunspot butterfly diagram. The semi-synthetic reconstruction was successful for both
weak and strong cycles (see their Fig.\ 13 and in particular Fig.\ 14, which shows a good match for cycle 14, 
 a weak cycle, and cycle 19, a strong cycle). This ability
to reconstruct (or forecast) both weak, medium, and strong cycles is very important and we shall come back to this point later in our own analysis.
\citet{2011A&A...528A..82J} also used those statistical correlations to go back in time, i.e.  to recreate the sunspot diagram from 1700 to 1874,
opening a very interesting research avenue now that some pre-1874 butterfly diagrams  have been recovered
from the full-disc
drawings of the Sun for the period from 1825 to 1867 prepared by Schwabe  
 \citep[see][]{2011IAUS..273..286A, 2011AN....332..805A, 2013MNRAS.433.3165A, 2015A&A...584A..73S, 2016A&A...592A.160L}. Other data is also
being recovered from historical drawings \citep[see][]{2008SoPh..247..399A, 2009SoPh..255..143A, 2009ApJ...700L.154U, 2015AN....336...53D}.

% new data in http://www.aip.de/Members/rarlt/sunspots

Later,  \citet{2016ApJ...823L..22C} made a prediction of the future of the entire cycle 24  (see their Fig.\ 1). It was based on
a surface flux transport model and data from synoptic magnetograms, which can be used to
predict the surface magnetic field in latitude and time, and as a consequence of the relationship between sunspots and magnetic fields can be
used to forecast the sunspot butterfly diagram itself.
\citet{0004-637X-792-1-12} also attempted to forecast the shape of cycle 25. They used the magnetic field data to extrapolate that the
first sunspots for cycle 25 will
appear in late 2019 (see their Fig.\ 17).

%% see papers before us
%% C:\Users\eurico\Downloads\aa16167-10.pdf
%% C:\Users\eurico\Downloads\apjl_823_2_L22.pdf

% put all references to all predictions
% http://ac.els-cdn.com/S1364682608003787/1-s2.0-S1364682608003787-main.pdf?_tid=d0b6a946-b1a4-11e5-a088-00000aab0f26&acdnat=1451775894_a19fac1fa490add01a5860bc6db7537c

 Overall, we believe that by including one further dimension, it  should be possible to choose between the multitude of forecasting methods.
In Section \ref{method} we describe the general method of spatial-temporal reconstruction using non-linear embeddings, and in particular the application
to a two-dimensional data set (one space, one time dimension). In Section \ref{parameters} we describe how the method's parameters can be auto-calibrated using
two statistical measures. In Section \ref{results} we apply the method to the sunspot area coverage latitude-time data set and finally in Section
\ref{conclusion} we conclude and draw on possible future research paths.

\section{Method: spatial-temporal reconstruction using non-linear embeddings}
\label{method}

Here we propose an approach which is drawn from the study of dynamical systems. Our approach is based on
the embedding reconstruction of local states from chaotic dynamical
systems theory applied to both the spatial and temporal components of the input series to forecast.
The reconstruction preserves the dynamics under smooth
coordinate transformations and the theorems by
\citet{key1503303m}, \citet{1981LNM...898..366T, 1981LNM...898..230M}, and \citet{1991JSP....65..579S}
guarantee the existence of the embedding.
The Whitney Embedding Theorem implies
that each state can be identified uniquely by a vector of $2n+1$ measurements, therefore reconstructing the phase space. The
Takens Embedding Theorem refines the approach to show that the reconstruction can be reached with a single measured quantity. Takens proved that
instead of $2n+1$ generic signals, the time-delayed versions of one generic signal is sufficient to recreate the $n$-dimensional
manifold. Similar theoretical results were obtained in \citet{zbMATH03744996} and a more empirical account \citep{PhysRevLett.45.712} was published around the same time.
However, the theorems
indicate an embedding dimension which is sufficient (but not
necessary) and is often too high for computational purposes.
In order to find more appropriate dimensions for computations
we use an approach that results from a refinement of the method described by
\citet{2000PhRvL..84.1890P} for the
reconstruction of spatial-temporal time series (STTS). They applied their method successfully to both
a spatial-temporal extension of the H\'enon map and to the Kuramoto-Sivashinsky (KS) equation. This method was later
also applied successfully to financial spatial-temporal series (yield curves) by \citet{Covas}, and by others to other data sets
\citep{PhysRevE.92.042910, Pan2008, BorstnikBracic2009, S021812740401045X, S0218127401003802}.

\subsection{The Parlitz-Merkwirth method}
\label{Parlitz-Merkwirthmethod}

We  now describe the method of Parlitz-Merwirth
\citep{2000PhRvL..84.1890P} of reconstructing local state data and how it can be applied to spatial-temporal data sets.
Let $n=1,...,N$ and $m=1,...,M$.
Consider a spatially extended time series ${\bf s}$ which can be
represented by a $N\times M$ matrix with components
$s^n_m \in \mathbb{R}$, which we call {\em states} of the STTS.
Consider a number $2I\in \mathbb{N}$ of spatial neighbours of a given
$s^n_m$ and a number $J\in \mathbb{N}$ of temporal past neighbours to $s^n_m$ (see Fig.\ \ref{forecasting_xfig}).
For each $s^n_m$, we define the {\em super-state} vector ${\bf x} (s^n_m)$ with components given by
$s^n_m$, its  $2I$ spatial neighbours, and its $J$ past temporal
neighbours, and with $K$ and $L$ being the spatial and temporal lags, respectively:
\begin{multline}
\label{embedding}
{\bf x}(s^n_m)=\{s^n_{m-I K }, ...,s^n_m, ..., s^n_{m+I K},
s^{n-L}_{m-I K},..., s^{n-L}_{m},..., s^{n-L}_{m+I K},
...\\ \cap
s^{n-J L}_{m-I K},..., s^{n-J L}_{m},..., s^{n-J L}_{m+I K} \}
\end{multline}
So the dimension of each ${\bf x}(s^n_m)$ is
\begin{equation}
d=(2I+1)(J+1).
\end{equation}
In their article,
Parlitz-Merkwirth use only rectangular
regions for the spatial-temporal neighbours of the centre element ${\bf x} (s^n_m)$ in
order to reconstruct ${\bf x}^i_j$. Other possible regions can be imagined, such as triangular regions
(designated by {\em lightcones}). In \citet{Covas} this triangular region approach was used with some success.
These triangular regions try to simulate the effect of the finiteness of information transmission across space, namely that information
cannot move faster than the speed of light.

Regarding boundary conditions, we follow  Parlitz-Merkwirth. Owing to the boundary
of the STTS, components of the local state vector ${\bf x} (s^n_m)$ in (\ref{embedding}) are not available when trying to construct states near
to the STTS boundary. This problem can be overcome by
extending the STTS in its spatial direction with numbers
$-c$, $-2c$, $-3c$, ... to the left and $+c$, $+2c$, $+3c$, ... to the right. The parameter $c>0$ has to be chosen
larger than the highest absolute value of the STTS. Using this
construction all states close to the boundary of the STTS
are located in different subspaces of the reconstruction
space.

Now, for each pair $(n, m)$, there is a one-to-one invertible map $f^{-1}$:
\begin{eqnarray}
f^{-1}: \mathbb{R} & \to & \mathbb{R}^d\\
s^n_m & \to & {\bf x}^n_m\equiv{\bf x} (s^n_m). \nonumber
\end{eqnarray}
We  now  approximate $f: \mathbb{R} \to \mathbb{R}^d$.

Take $N_{training}$ time consecutive states
$s^n_m$ of ${\bf s}$. With these states we form a training set $\mathbb{A}$ of super-states ${\bf x}^n_m$.
We  reconstruct a given super-state ${\bf x}^n_m\in \mathbb{A}$ by using its
closest past neighbours on $\mathbb{A}$, separated in time by $\tau \in
\mathbb{N}$.

We  then approximate $f$ by some unknown
function $F: \mathbb{R}^d\to \mathbb{R}$ such that
\begin{equation}
F(x^n_m)=s^{n+\tau}_m.
\end{equation}
There are several ways to do this. In their article,
Parlitz-Merkwirth propose the following method: take a ${\bf x}^n_m$. Find the nearest
neighbour to ${\bf x}^n_m$ on $\mathbb{A}$, say ${\bf x}^i_j$, in the Euclidean norm. Now,
$s^{i+\tau}_j$, which is known a priori, will be an approximation
$p^{n+\tau}_m$ for
$s^{n+\tau}_m$, i.e.
\begin{equation}
F(x^i_j)\equiv s^{i+\tau}_j \approx s^{n+\tau}_m,
\end{equation}
where $x^{i}_j$ is the nearest neighbour of $x^n_m$.
We note that forecasts
over longer periods of time $(\tau >1)$ can be calculated
as a single large step $\tau$ or iteratively by concatenating steps
with $\tau = 1$.

We  introduce another possible modification here with respect to the
original method
by considering the $n$th nearest
neighbouring  super-state to ${\bf x}^n_m$
and then averaging the $n$ values of $s$ obtained in this way in order
to get $s^{n+\tau}_m$. This neighbourhood averaging  also carries
some weight according to the Euclidean distance to the central
super-state $x^n_m$. In \citet{Covas} this averaging approach was used with some success.

%Finally, we shall introduce a smoothing method after we get the data from the
%above (modified) Parlitz-Merkwirth procedure. For the smoothing we shall use polynomial
%least squares method (the polynomial order will depend on the case). Again in Covas and Mena \citep{Covas} this smoothing approach was used with %some success.

The embedding theorems do not state how to choose
the space and time delays of the embedding.
This can be done using the
notion of {\em average mutual information}, which
has been used widely in the past (see e.g.\ \citet{abarbanel1997analysis, opac-b1092652}).
This  gives us an estimate for the values of the spatial and temporal delays $K$ and
$L$, which can then be used to determine $I$ and $J$ and therefore the embedding dimension.
Mutual information estimates how measurements of $s^i_j$
at time $i$
are connected to measurements of $s^{i+L}_j$ at time $i+L$.

After we calculate the (spatial and temporal) lags $K$ and $L$, we can proceed to determine the embedding parameters $I$ and $J$, for which we  use
the {\em method of false neighbour} detection proposed by
\citet{1992PhRvA..45.3403K} and described in detail in
\citet{1996PhT....49k..86A} and \citet{abarbanel1997analysis}. This approach calculates the number of neighbours of a point and how that number
changes with increasing embedding dimension. Below the theoretical embedding dimension, many of these neighbours will be false, due to projection,
but at  a higher embedding dimension all neighbours are real. The advantage of using this auto-calibration as opposed to
choosing those parameters based on the best match to the future observed data set is that we avoid any bias. This way the calibration
is fully based on the training set and nothing else and is an unbiased approach.

\section{Parameter estimation}
\label{parameters}

We calibrate our parameters $I$, $J$, $K$, $L$ in equation (\ref{embedding})
using the average mutual information first minimum \citep[see][]{Fraser86} for the $K$ and $L$ (spatial and temporal) lags and the false neighbours method of
finding the optimal embedding (spatial and temporal) dimensions $I$ and $J$ \citep[see][and references therein]{2000PhRvL..84.1890P}.
In other words, we take slices in space (or in time), calculate the one-dimensional mutual information and percentage of false neighbours,
and then average the values. We can then calculate an estimate for the best  spatial (and temporal) lags and embeddings.
%Notice
%that these 4 parameters are not arbitrary or free parameters since the mutual information first minimum approach and the false neighbours
%method should give the optimal embedding. This is a crucial and very important part of this method, as it is ``free'' from parameters
%that are calculated from the future states we want to forecast. In this way we believe there is no ``bias'' in this calibration.

The first two parameters to calibrate are therefore
the $K$ and $L$ (spatial and temporal) lags, inferred by finding the first minimum of mutual information.
The mutual information is calculated as follows. Let $s^{i}$ be a one-dimensional data set and $s^{i+L}$ the related $L$-lagged data set.
We note that we have to truncate the set
$s^{i}$ by $L$ points in order to take into account its size when lagging it.
Given a measurement $s^i$, the amount of information $I(L)$ is the number of bits
on $s^{i+L}$, on average, that can be predicted, and is calculated as 
\begin{equation}
\label{mutualinfotime}
I(L)=\sum_{i}\sum_{i+L} P(s^i,s^{i+L})\log_2\frac{P(s^i,s^{i+L})}
{P(s^i)P(s^{i+L})},
\end{equation}
where $P(s^i)$ is the probability of finding a time series value in the $i$-th interval and
$P(s^i,s^{i+L})$ is the joint probability of measuring $s^i$ and $s^{i+L}$.
%To calculate the probabilities, one has to choose a bin size.
%There are main
%We use the Freedman-Diaconis rule \citep{6360487}
%\begin{equation}
%\label{FreedmanDiaconisRule}
%{\textnormal{Bin Size}}=2\, {\textnormal{IQR}}(x) \, n^{-1/3},
%\end{equation}
%where ${\textnormal{IQR}}(x)$ is the interquantile range of the data and $n$ is the number of observations in the sample $x$.

% see http://stackoverflow.com/questions/14683467/finding-the-first-and-third-quartiles
% see https://en.wikipedia.org/wiki/Freedman%E2%80%93Diaconis_rule
% see https://en.wikipedia.org/wiki/Interquartile_range
% IQR = Q3 -  Q1

Following on from Fraser and Swinney's article (\citealt{Fraser86}; see also \citealt{1992PhRvA..45.7058M, 1993RvMP...65.1331A}), the maximum 
unpredictability coincides with the minimum  predictability,
i.e. at a minimum in the mutual information. As chaotic time series diverge exponentially, due to one or more
positive Lyapunov exponents \citep{9814277657},
the first minimum of $I(L)$ -- rather than some
subsequent minimum -- should probably be chosen for the time lag $L$ to sample the data.

In order to calculate the two lags for a two-dimensional set, we take slices in both space and in time, calculate the mutual information, average it, and then
find the first minimum to obtain the $K$ and $L$ (spatial and temporal) lags.
Once the lags are obtained, the next step in the calibration or parameter estimation is to calculate the minimum embedding dimension, or equivalently,
the number of spatial and temporal neighbours to use in the phase space reconstruction. To do this, we use the method of false neighbours, as described in
\citet{1992PhRvA..45.3403K, 1992PhRvA..45.7058M} and \citet{1993RvMP...65.1331A}. This approach determines, directly from the data, when the apparent close neighbours
or crossing of orbits have been eliminated by virtue of projecting the full orbit in a too low dimensional embedding phase space.
The implementation by \citet{1992PhRvA..45.3403K} is as follows. If we have an embedding in $d$ dimensions  for a time series $s(n)$,  and we define the distance
of a phase space vector ${\bf x}(n)$ to its $r$th nearest neighbour ${\bf x}^{(r)}(n)$ by the square of the Euclidian distance,
\begin{equation}
R_d^2(n,r) = \sum_{i=0}^{d-1} {\left[ s ( n + i L) - s^{(r)} (n + i L) \right]^2},
\end{equation}
then as we increase the dimension to $d+1$, we add a new coordinate $s ( n + d L) $ to the ${\bf x}(n)$ vector.
So the new distance in this $d+1$ space is
\begin{equation}
R_{d+1}^2(n,r) = R_{d}^2(n,r) + {\left[ s ( n + d L) - s^{(r)} (n + d L) \right]^2}.
\end{equation}
If there are two false neighbours,  as we increase the dimension from $d$ to $d+1$, it is expected that this distance will increase substantially. \citeauthor{1992PhRvA..45.3403K}
define a criterion for this by designating a false neighbour as one that has
\begin{equation}
\sqrt{\frac{R_{d+1}^2(n,r) - R_{d}^2(n,r)}{R_{d}^2(n,r)}}=
\frac{\left| s ( n + d L) - s^{(r)} (n + d L) \right|}{R_{d}(n,r)}>R_{tol},
\end{equation}
where $R_{tol}$ is some arbitrary threshold. \citeauthor{1992PhRvA..45.3403K} recommend  using a threshold of around 10 and we  use the same value later.
We record and output the percentage of presumed false neighbours as a function of the embedding dimension $d$. Since we have a spatial-temporal signal,
again we slice in space and in time, calculate the percentage of false neighbours, and then average over space and time as a function of $J$ and $I$,
respectively.

This slicing in space and time is obviously an arbitrary choice. However, there is not -- as far as we are aware -- a
mathematical generalisation for the calibration
of the lag and dimension parameters for a multi-dimensional data series. We have not attempted in this article to explore other approaches, although
we recognise that studying the stability of the embedding, and of the forecasting, as a function of the slicing and averaging methodology is  an interesting
research topic. A mathematical theory of mutual information in higher dimensions and a theoretically justifiable method for choosing the optimal lags and embedding dimensions in higher dimensions are sorely needed.

\section{Results}
\label{results}

\subsection{ Data}

We analyse data for sunspot area coverage
from 1874 to 2015\footnote{We use the
data set publicly available in \url{http://solarscience.msfc.nasa.gov/greenwch/bflydata.txt } \citep{butterfly}.}. The data we used starts
at Carrington Rotation\footnote{Since the solar rotation at the surface varies with latitude and time,
any method of comparing positions on the solar surface over a period of time is obviously subjective.
Solar rotation is arbitrarily taken to be 27.2752316 days
for the purpose of Carrington rotations. Each rotation of the Sun is given a unique
number called the Carrington Rotation Number, starting from November 9, 1853.} number 275 and finishes at 2162.
The data is made of blocks of 50 data points per Carrington Rotation,
each point representing a latitudinal bin, all of which are distributed uniformly in $sin(\textnormal{latitude})$.
The data points represent
the area of that time/latitude `rectangle' that is occupied by sunspot(s) in units of millionths of a hemisphere\footnote{There are several technicalities in
correcting and calibrating the source sunspot area data across the entire time period of three centuries. We direct the reader to
\url{http://solar-b.msfc.nasa.gov/ssl/PAD/solar/greenwch.shtml} for these details.
}.

We take as a `training set' the data from  the year 1874 (i.e.\ the first 1646 Carrington Rotations) to approximately  1997; that is, we take as a training set the data for Carrington Rotations 275 to 1920 inclusive. We then try to
forecast the sunspot area butterfly diagram from Carrington Rotation 1921 to 2162 (the latter representing approximately the year 2015); that is, we use 1646 time slices
(approximately 122.92 years)
to forecast the next 242 time slices (approximately 18.07 years). The training set represents approximately 12 solar cycles (cycle 11 to 22),
while the `forecasting set'
represents approximately 1.5 cycles (cycle 23 and half of cycle 24).
The entire data set, including the training set and the forecasting set, is a grid $A^i_j=A(i,j)$, with $i=1888$ and $j=50$. The training set is a grid $A(1646,50)$.
The area values,
in units of millionths of a hemisphere, range from 0 when there are no sunspots to the maximum value of 2580. Of the total $i\times j= 94,400$ data points, approximately
$27.54\%$ are non-zero, i.e.\ they represent a sunspot area occupying that time/latitude rectangle. For the entire data set and taking into
account the points for which $A(i,j)=0$ the average is $\langle A(i,j) \rangle \approx 16.77$.

The distribution of area covered by sunspots as a function of time and latitude is by no means uniform, as can be seen in  Table \ref{histogram}. 
This distribution will later influence the way we colour the real and forecasted sunspot diagram (see
Figs.\ \ref{Spatiotemporal_Forecasting_Sunspots_LastCycle} and \ref{Spatiotemporal_Forecasting_Sunspots_LastCycle_Amplification}).
\begin{table}[htb]
\centering
\caption{Sunspot area coverage (in millionths of a hemisphere) showing the number  of data points and the  percentage  of the total
within certain area bins.}
\label{histogram}
\begin{tabular}{rrrrr}
{\bf Area coverage bin} & {\bf Count } & {\bf Percentage of total }     &  &  \\
0  & 68398 & 72.46\% &  &  \\
1    & 5790  & 6.13\%  &  &  \\
2    & 1189  & 1.26\%  &  &  \\
5    & 2105  & 2.23\%  &  &  \\
10   & 2183  & 2.31\%  &  &  \\
20   & 2555  & 2.71\%  &  &  \\
50   & 4045  & 4.28\%  &  &  \\
100  & 3362  & 3.56\%  &  &  \\
$\infty$ & 4773  & 5.06\%  &  &
\end{tabular}
\end{table}

\subsection{Calibration of $I$, $J$, $K$, and $L$}

We first calibrate the $L$ parameter. We slice the training set into 50 (latitudinal) pure time series, and then calculate the mutual information
as a function of $L$ for each  of these slices. Then we average over the 50 slices, showing the average of the mutual information in order to make a decision on
the optimal $L$-temporal lag.
\begin{figure}[htp]
\resizebox{\hsize}{!}{\includegraphics{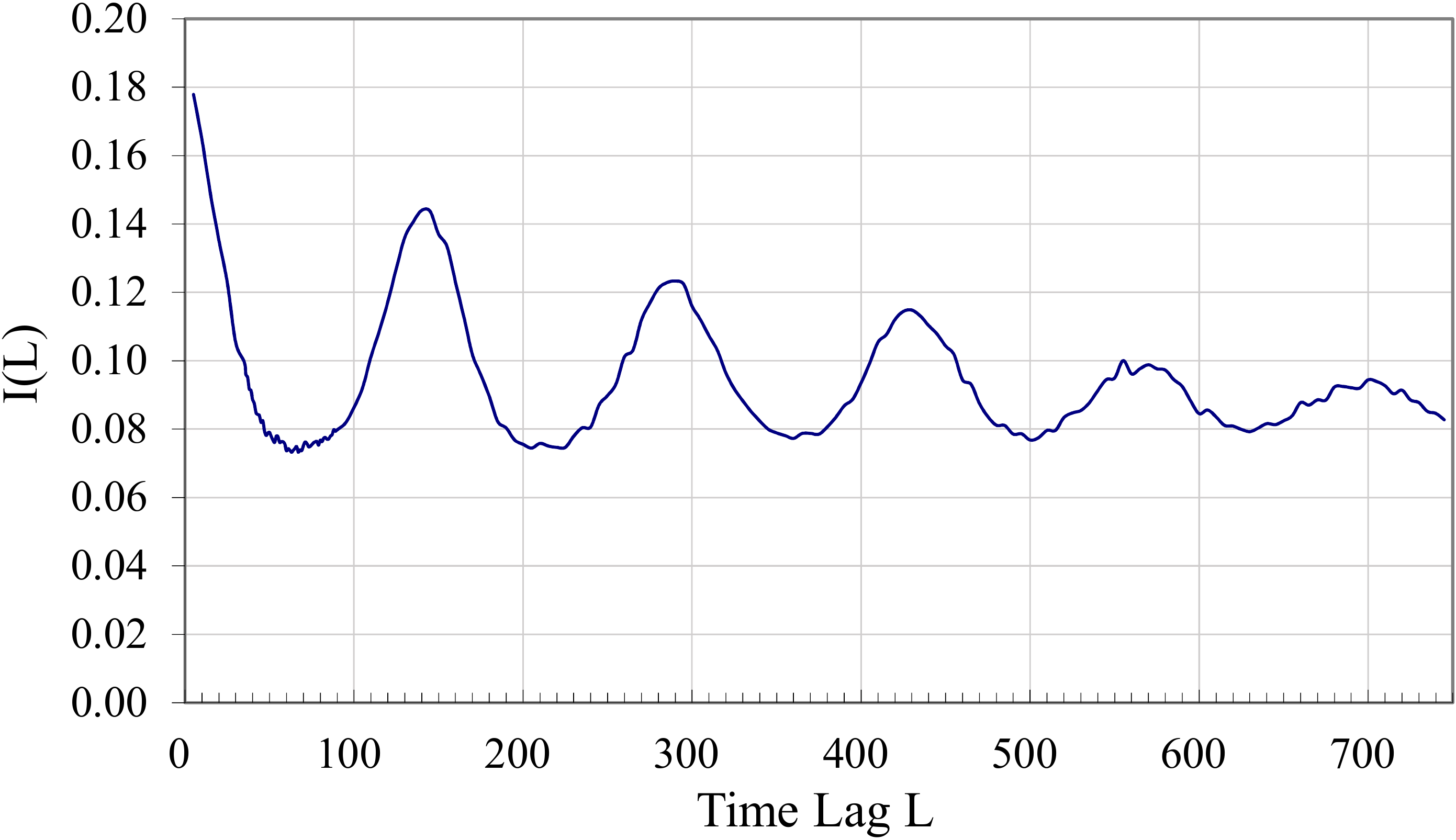}}
\caption{Average mutual information as a function of time lag $L$, $I(L)$. Although it is not clear
what the exact first minimum is, due to data noise, it is close to or around $L=70$ months, which we will take as the first minimum.}
\label{Mutual_Information_function_L}
\end{figure}
We can see in Fig.\ \ref{Mutual_Information_function_L} the usual mutual information shape for a time series, with a collection of successive local minima. We  use the first minimum \citep{Fraser86,opac-b1092652,9814277657} as an indicator
of the best value for the $L$-temporal lag parameter.
There is clearly some noise in the figure;  we  note that, theoretically, the method assumes we are dealing with infinite, noise-free trajectories and there is no guarantee it will work for real physical data \citep{1999PhR...308....1S}.
The first minimum is around $L=70$, corresponding to around 5.22 years, and we  use this value hereafter.
We also calculated the minima for the pure time series $A(t)=\sum_{\textnormal{latitude}} A(t,\textnormal{latitute})$,
representing the total sunspot area across the entire solar disc,  and we have found it to be $L=58$,
corresponding to around 4.33 years or around 52 months.  \citet{zhoufeng}, who have analysed the smoothed average sunspot count (as opposed to the
sunspot area, but related), have arrived at a time-lag value of 38 months, while  \citet{1990SoPh..126..407K} and \citet{1991JGR....96.1705M} 
have arrived at
a value of 2-5 years and 10 months respectively. Others   have found similar values \citep{2001A&A...377..312S,
2006A&A...449..379L,JiangS11, 2016AJ....151....2D}. For smaller daily sunspot count time series \citet{2001A&A...379..611J} arrived
at optimum time delay of 8-12 days, but we believe this is due to the effect of having both daily points and less than one solar cycle (they used
around 3346 days of daily data).

\begin{figure}[htp]
\resizebox{\hsize}{!}{\includegraphics{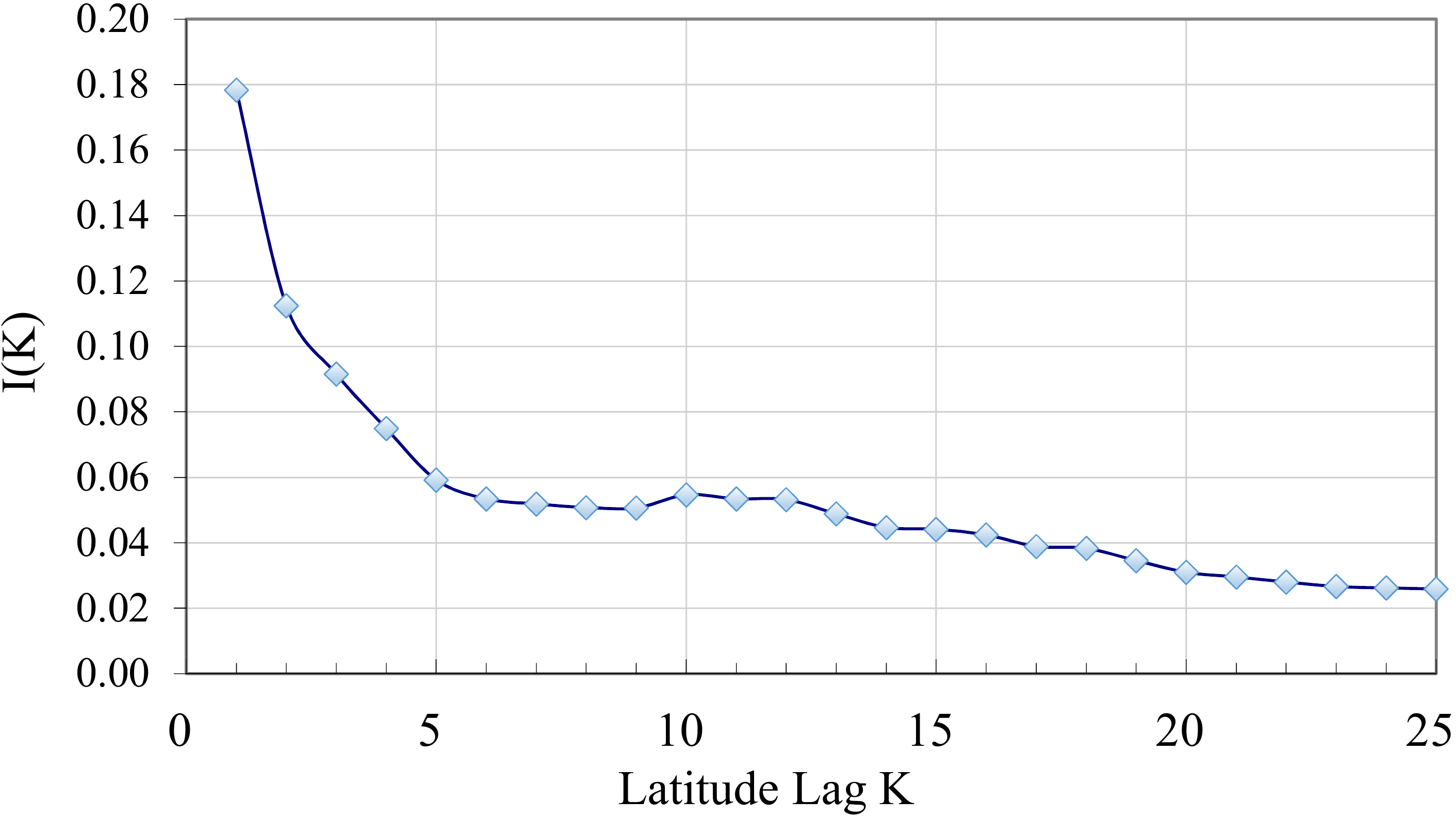}}
\caption{Average mutual information as a function of latitude grid spacing lag $K$, $I(K)$. The first minimum seems to indicate optimum
latitude grid spacing of around $K=9$.}
\label{Mutual_Information_function_K}
\end{figure}

%\begin{figure}
% \resizebox{\hsize}{!}{\includegraphics{Mutual_Information_function_L_amplified.pdf}}
% \caption{Zoom in average mutual information as a function of time lag $L$, $I(L)$. Although it is not clear
% what the exact first minimum is, it is close or around $L=70$ months.}
% \label{Mutual_Information_function_L_amplified}
%\end{figure}
In a similar way we shall calibrate the $K$ parameter. We slice the training set into 1646 (latitudinal) pure time series, and then calculate the mutual information
as a function of $K$ for each one of those slices. Then we average the resulting mutual information across the 1646 slices, showing the average to make a decision on
the optimal $K$-spatial lag (Fig.\ \ref{Mutual_Information_function_K}). The first minimum is around $K=9$ and we  use that value thereafter.

\begin{figure}[htp]
\resizebox{\hsize}{!}{\includegraphics{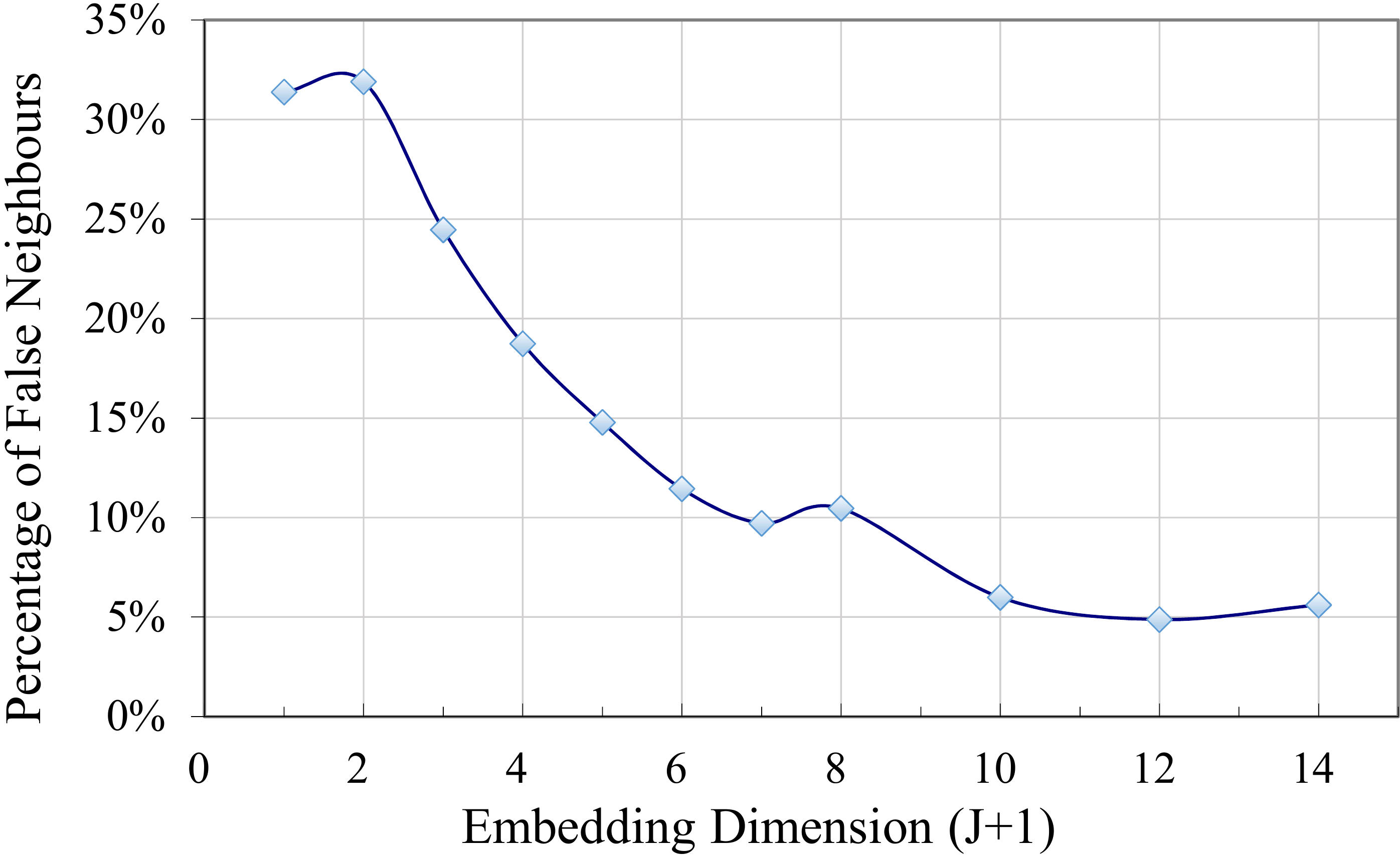}}
\caption{Percentage of false nearest neighbours averaged over all latitudes. We take the minimum
embedding dimension to be the one for which the percentage is first 10\% or less.
The data indicates that  $J+1 \approx 7 \implies J = 6$.}
\label{Percentage_False_Nearest_Neighbors_J}
\end{figure}

\begin{figure}[htp]
\resizebox{\hsize}{!}{\includegraphics{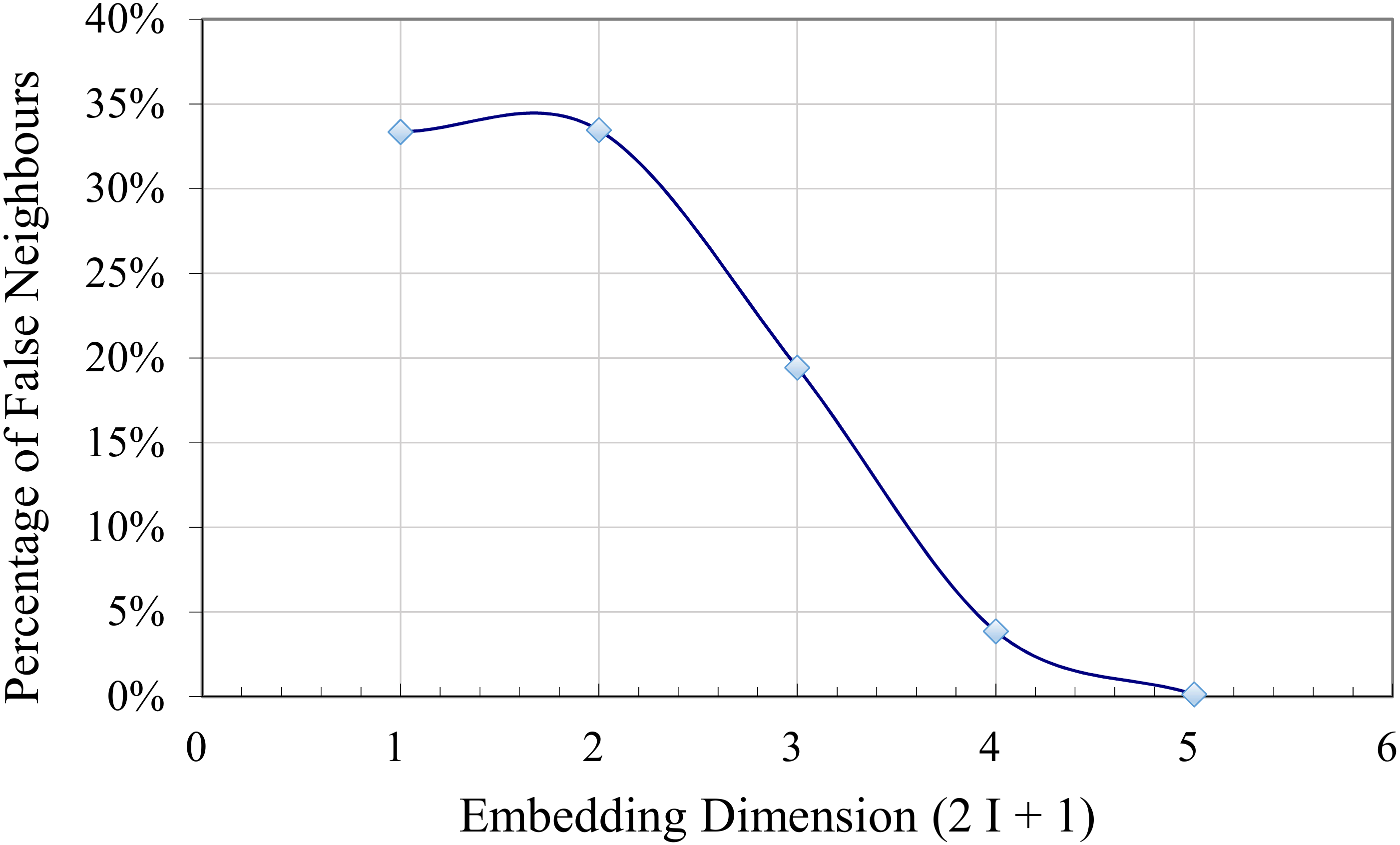}}
\caption{Percentage of false nearest neighbours averaged over all time slices. We take the minimum
embedding dimension to be the one for which the percentage is first 10\% or less. The data indicates that  $2I+1 \approx 4 \implies I = 2$
(we truncate $I$ as we want the embedding area to be symmetric with respect to the centre element ${\bf x} (s^n_m)$).}
\label{Percentage_False_Nearest_Neighbors_I}
\end{figure}

Now that we have $K=9 $ and $ L=70$ we can use them to create successive higher dimensional embeddings in both space and time to
calculate the $I$ and $J$ optimal parameters. We take the same approach that we took for the calibration of the lags,  and
average the percentage of false neighbours over spatial/temporal dimensions. As indicated in Section \ref{parameters}, we use $R_{tol}=10$ although we tested the
stability of the implied minimum embedding dimension to several values of $R_{tol}$. This can be seen in Figs.\ \ref{Percentage_False_Nearest_Neighbors_J} and \ref{Percentage_False_Nearest_Neighbors_I} as a function of $J$ and $I$, respectively.

Given the shape of the percentage of false nearest neighbours is affected by noise and the finite aspect of the time series, we take
the assumption that the embedding dimension is found whenever the percentage is below 10\%. The figures imply that
$I=2$, $J=6$ and we  use these values to attempt to forecast the sunspot diagram in the next section.

\subsection{Forecast}
\label{originalforecast}

\begin{figure*}[htp]
\resizebox{\hsize}{!}{\includegraphics{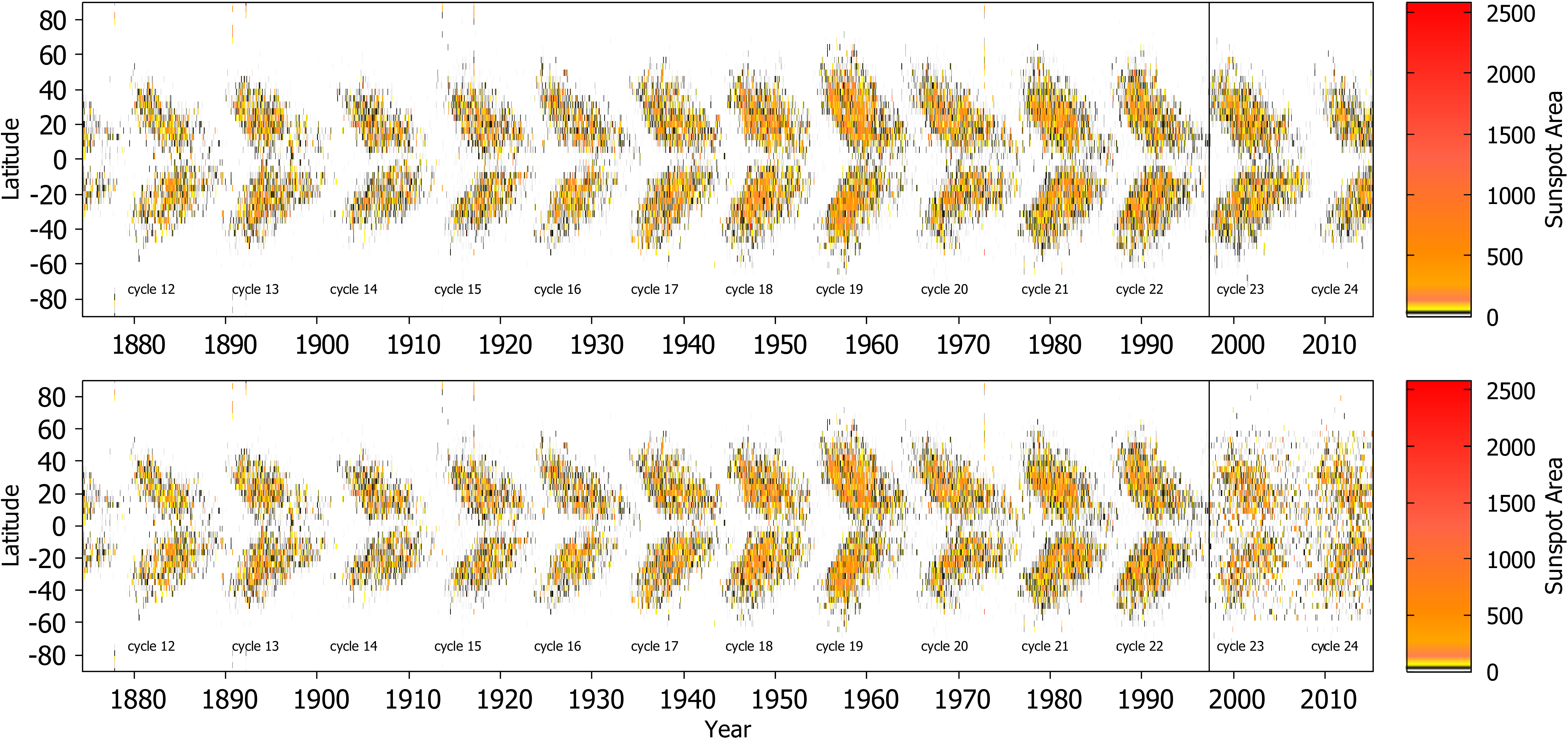}}
\caption{Spatial-temporal forecast of the last two cycles using the calibrated parameters $I=2$, $J=6$, $K=9$, and $L=70$. The top panel
is the full data set, including the training set and the actual observed future data set. The bottom panel is the training set plus
the forecasted set.
The boundary between the training set and the forecast set is marked by a black vertical line in each plot. The main
features of the butterfly diagram, namely the amplitude of the cycle and the migration to the equator, are both present, even if the forecast
is far from perfect. The colour scheme used tries to follow the distribution of covered area as suggest by the data in Table \ref{histogram}.}
\label{Spatiotemporal_Forecasting_Sunspots_LastCycle}
\end{figure*}

\begin{figure}[htp]
\resizebox{\hsize}{!}{\includegraphics{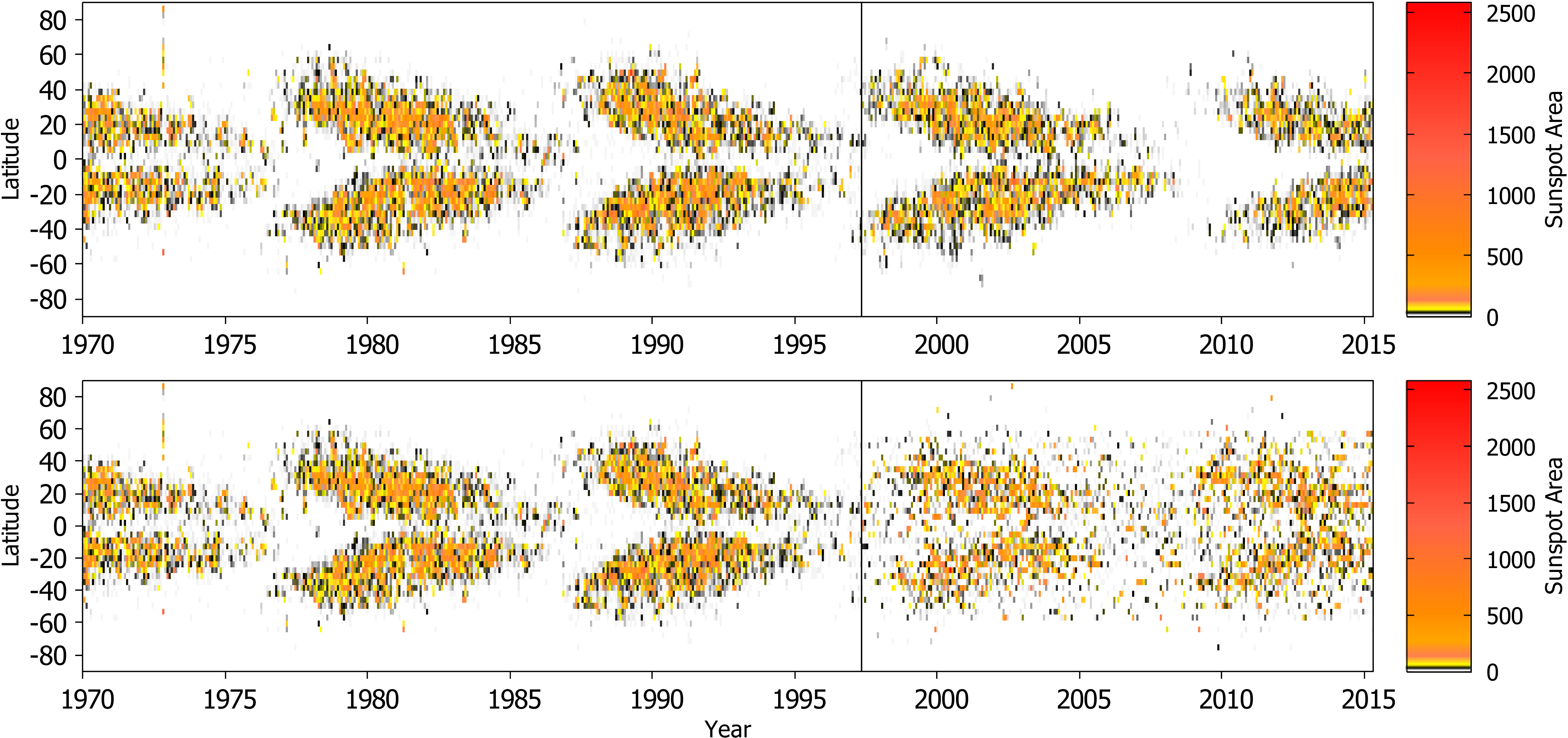}}
\caption{Amplification of the spatial-temporal forecast of the last two cycles using the calibrated parameters $I=2$, $J=6$, $K=9$, and $L=70$.
The top panel
is the full data set, including the training set and the actual observed future data set. The bottom panel is the training set plus
the forecasted set.
The boundary between the training set and the forecast set is marked by a black vertical line on each plot.}
\label{Spatiotemporal_Forecasting_Sunspots_LastCycle_Amplification}
\end{figure}

Using the best parameter calibration, we have $I=2$, $J=6$, $K=9$, and $L=70$, and we  now attempt to forecast approximately 1.5 cycles (cycle 23 and half of cycle 24).
We forecast one data slice at a time, i.e.\ one Carrington rotation at a time, then concatenate, then use the same calibrated
parameters to forecast the next, and so on. A total of 242 data slices are forecast in this way. The results are shown in
Fig.\ \ref{Spatiotemporal_Forecasting_Sunspots_LastCycle} and the amplification in
Fig.\ \ref{Spatiotemporal_Forecasting_Sunspots_LastCycle_Amplification}.

The results show that we can reproduce the two main
features of the butterfly diagram; i.e. the amplitude of the cycle and the migration to the equator are both present, even if the forecast
is far from perfect. There seems to be a dispersion of points in the forecast. There are also some blips, probably caused by the noise
level, which may lead the algorithm to fail badly. Still overall, it is a  reasonable result, considering  that this particular
method is generic as it does not need to know the underlying physics of the system.
We also attempted to do the forecasting using the concept of lightcones, i.e.\ restricting the embedding vector to an isosceles triangle of data points
on the original super-state vector ${\bf x}(s^n_m)$ in (\ref{embedding}), but this did not improve the forecast. We also attempted to use the weighted
averaging as described in Section (\ref{Parlitz-Merkwirthmethod}), but again saw no noticeable improvement.  This is in contrast
with the results in \citet{Covas}, but we have not found an explanation for this behaviour. Still by not using any modification to
the auto-calibrated method, we ensure that no bias is introduced when choosing the details of the approach.

\begin{figure}
\resizebox{\hsize}{!}{\includegraphics{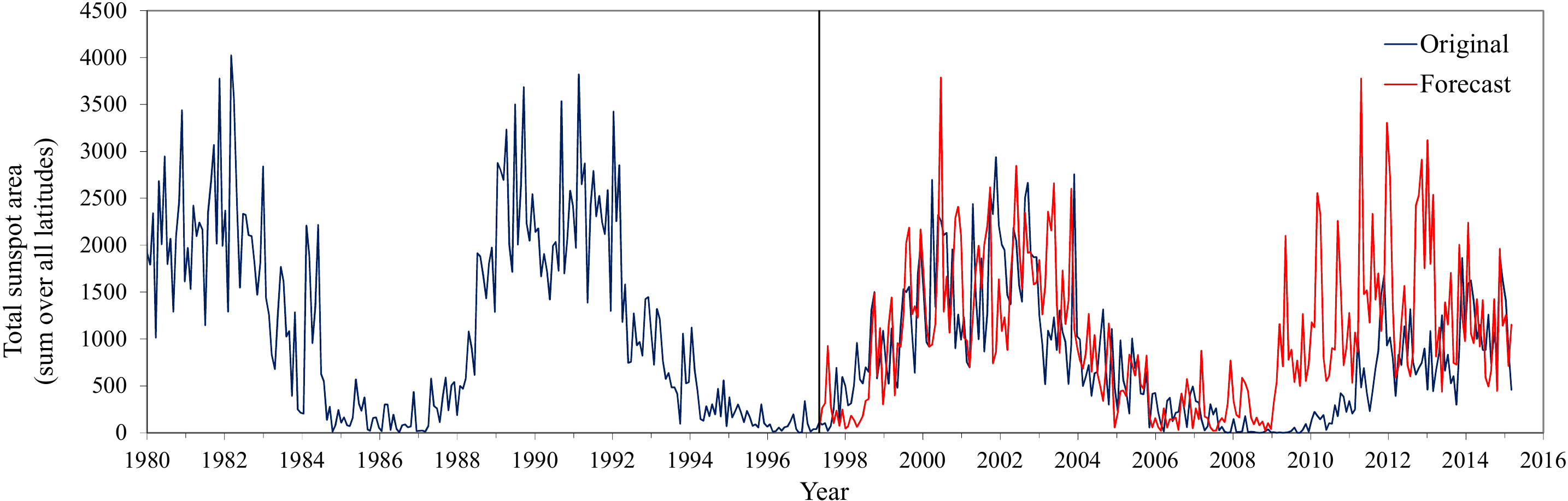}}
\caption{Overall sunspot area (sum over all latitudes) $A(t)$ and the forecast for the two last cycles using parameters
$I=2$, $J=6$, $K=9$ and $ L=70$. The black line marks the start of the
forecast.}
\label{Forecast_Sunspot_Area_Amplification}
\end{figure}

 Figure\ \ref{Spatiotemporal_Forecasting_Sunspots_LastCycle} and the amplification in
Fig.\ \ref{Spatiotemporal_Forecasting_Sunspots_LastCycle_Amplification} show that the method works qualitatively well for both
the amplitude of the cycle and the migration to the equator. Overall, the shape and angle of the latitudinal bands of the butterfly diagram
seem to be preserved by the forecast. In order to
put this conclusion on firmer ground, we calculated the overall sunspot area (sum over latitude) for the forecast and compared it with the
observed one. These results are depicted in Fig.\ \ref{Forecast_Sunspot_Area_Amplification}, which  shows the total sunspot area, and both the original training set
(and observed future set)
and the forecasted set using the same parameters $I=2$, $J=6$, $K=9$, and $L=70$. Although there is quite a lot of noise (in both sets), it shows
that the method seems to work at least qualitatively not only in space and time, but also on aggregated metrics such as the sum of the area over the latitude.
The question of whether  it really has  predictive power is  addressed in Section \ref{Effectiveness}. 

%\begin{figure}
% \resizebox{\hsize}{!}{\includegraphics{Forecast_Sunspot_Area-crop.pdf}}
% \caption{Overall sunspot area (sum over latitudes) and the forecast for the two last cycles using parameters $I=2$ , $J=6$ , $K=9 $ and $ L=70$. %The black
% line marks the start of the
% forecast.}
% \label{Forecast_Sunspot_Area}
%\end{figure}

\begin{figure}
\resizebox{\hsize}{!}{\includegraphics{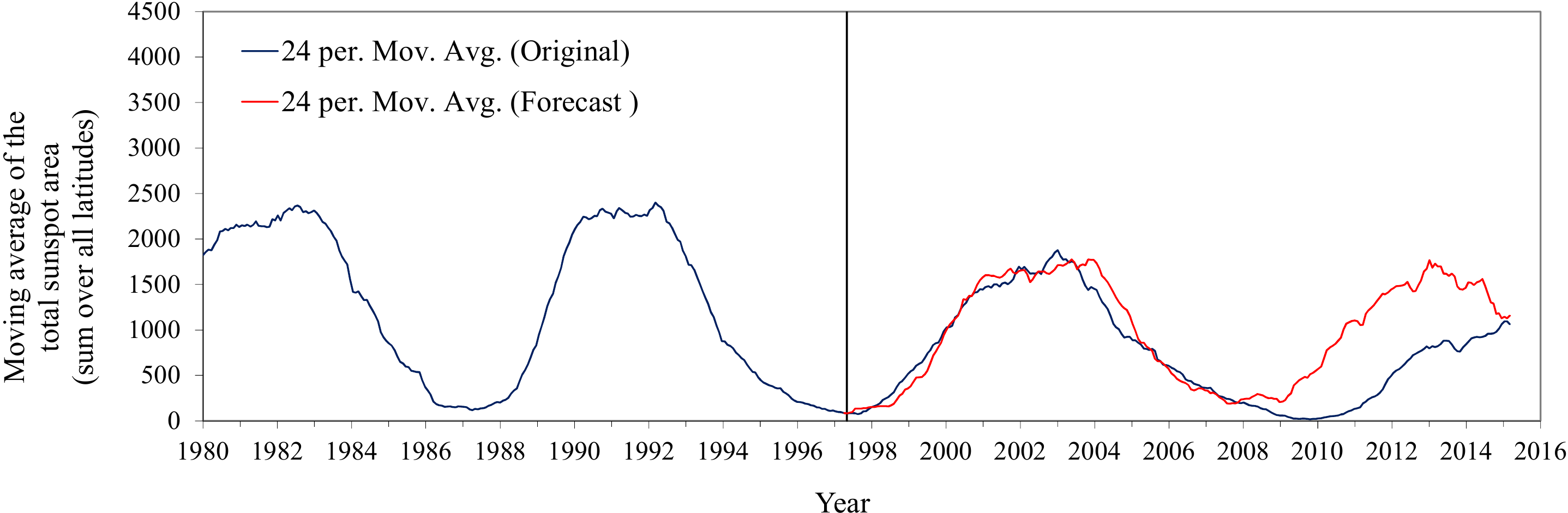}}
\caption{Moving average (24-point average) total sunspot area (sum over all latitudes) $\overline{A(t)}$ and the forecast for the two last cycles using parameters $I=2$, $J=6$, $K=9 $ and $ L=70$. The black
line marks the start of the
forecast.}
\label{Forecast_Sunspot_Area_RunningAverage_Amplification}
\end{figure}

We also calculated the 24-point moving average
\begin{equation}
\overline{A(t)} \doteq \left\langle \sum_{\textnormal{latitude}} A(t,\textnormal{latitude}) \right\rangle_{24}
\end{equation}
to show the overall smoothed total sunspot area cycle against the original cycle. The
moving average was taken backwards in time and the total sunspot area is the sum over all latitudes of $A(t,\textnormal{latitude})$. The results are depicted in
 Fig.\ \ref{Forecast_Sunspot_Area_RunningAverage_Amplification}, which shows that the method is quite good at reproducing
the first cycle (cycle 23), but -- not surprisingly --  that it starts to fail when forecasting the next cycle (cycle 24). This could indicate that the method
can only reproduce the cycle a few years ahead or -- even  worse -- that it fails badly for weak cycles like cycle 24. 
%Still the approach seems to be able to forecast quite well the entire shape of cycle 24.

%\begin{figure}
% \resizebox{\hsize}{!}{\includegraphics{Forecast_Sunspot_Area_RunningAverage.pdf}}
% \caption{Running average (24 months average) sunspot area (sum over latitudes) and the forecast for the two last cycles using parameters $I=2$ , $J=6$ , $K=9 $ and $ L=70$. The black
% line marks the start of the
% forecast.}
% \label{Forecast_Sunspot_Area_RunningAverage}
%\end{figure}

%The other significant feature of the butterfly diagram besides its amplitude modulation, is the migration to the equator. We can quantity this
%more precisely by calculating the average weighed latitude of the sunspot coverage, that is to calculate the following quantity:
%\begin{equation}
%\langle \theta^{\pm} \rangle (t) = \sum_{\theta^{\pm}} \theta \, A(t,\theta) \diff\theta,
%\end{equation}

%\begin{figure}
% \resizebox{\hsize}{!}{\includegraphics{Forecast_Sunspot_Area_AverageRunningAverageLatitude-crop.pdf}}
% \caption{Running average (24 months average) sunspot area average latitude and the forecast for the two last cycles using parameters $I=2$ , $J=6$ %, $K=9 $ and $ L=70$. The black
% line marks the start of the
% forecast.}
% \label{Forecast_Sunspot_Area_AverageRunningAverageLatitude}
%\end{figure}

\subsection{Structural similarity}

The method used here has the advantage of reproducing the entire spatial-temporal features of the sunspot butterfly diagram. However, it is
not easy to quantitatively estimate when we have found a good forecast. The human brain is able to look at the observed butterfly diagram
and the forecasted one in Fig.\ \ref{Spatiotemporal_Forecasting_Sunspots_LastCycle} and assess almost instantaneously whether it is a good or a bad forecast.
However, using numerical quantities such as the root-mean-square error (RMSE) may not be the best way to quantify the goodness of a forecast.
Although the RMSE has the advantage of being parameter free and cheap to compute, relatively simple, and memoryless, the RMSE can be
evaluated at each sample independently of the other samples; it has the disadvantage of
not being able to imitate the human perception of image similarity. In the particular case of the butterfly diagram, what we are after
is to closely match the overall butterfly wing shape, the migration to the equator, and the overall amplitude (width and intensity).

In order to assess the similarity of the forecast with the actual sunspot butterfly diagram we 
use the concept of structural similarity $\textnormal{SSIM}(x,y)$ introduced by \citet{Wang04imagequality},
\begin{equation}
\label{SSIM}
\textnormal{SSIM}(x,y) = \frac{(2 \mu_x\mu_y +c_1)(2\sigma_{xy}+c_2)}{(\mu_x^2 +\mu_y^2+c_1)(\sigma_x^2+\sigma_y^2+c_2)},
\end{equation}
where $\mu_x$ is the average of $x$, $\mu_y$ is the average of $y$, $\sigma_x^2$ is the variance of $x$, $\sigma_y^2$ is the variance of $y$,
$\sigma_{xy}$ is the covariance of $x$ and $y$, and $c_1$ and $c_2$ are constants implicit in the structural similarity method (for details see \citealt{Wang04imagequality,2012ITIP...21.1488B}).
The SSIM index is a method for calculating the perceived quality of digital images and videos.  It 
allows two images to be compared and provides a value of their similarity. 
 The SSIM index is a decimal value between -1 and 1;  a value of 1 is only attained in the case of two identical sets of data.
The SSIM index is designed to improve on traditional methods such RMSE, which have proven to be inconsistent with human visual perception\footnote{A particularly striking visual demonstration of the advantage of using SSIM over the RMSE index is shown in
\url{https://ece.uwaterloo.ca/~z70wang/research/ssim/\#test }.
}.

\begin{figure}[htp]
\resizebox{\hsize}{!}{\includegraphics{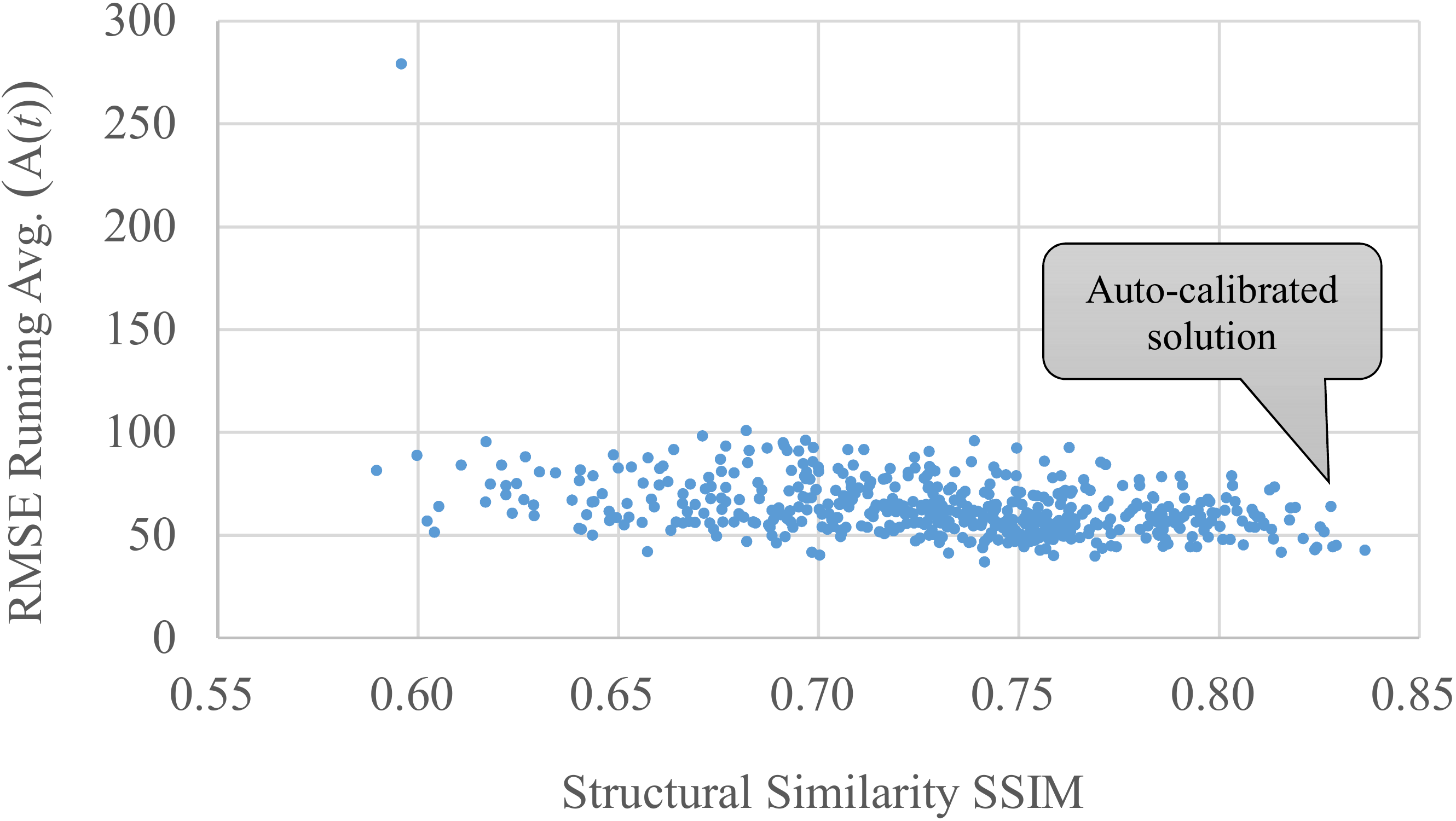}}
\caption{RMSE of the error (differences) between the observed and forecasted 24-point moving average of the total sunspot area
$\overline{A(t)}$
against the structural similarity SSIM for random choices
of the forecasting algorithm's input parameters. The higher the SSIM, the closer the forecast is to the observed sunspot butterfly diagram. This shows that similar quality forecasts,
at least in terms of structural similarity, can have quite different $\textnormal{RMSE}\left(\overline{A(t)}\right)$.}
\label{StructuralSimilarity_versus_SumSquaresDifferenceRunningAverageAreaSunspotsFunctionTime}
\end{figure}

We use the SSIM index to measure the similarity of the forecast and the original sunspot cycle. We calculate the index using the $242\times 50$ points that are forecasted,
i.e. the forecast over almost two cycles from Carrington rotation 1921. We first try to show
how the RMSE metric is a bad indicator of similarity by showing the RMSE versus the SSIM for a random Monte Carlo generated
collection of forecasts. We take random samples of $I=1,2$; $J=4,5,6$; $K=7,8,9$; and $L=60,\ldots,80$, together with the use of lightcones or not,
and calculate the RMSE of the error (differences) between the observed and forecasted 24-point moving average of the total sunspot area
$\textnormal{RMSE}(\overline{A(t)})$. This is depicted in Fig.\ \ref{StructuralSimilarity_versus_SumSquaresDifferenceRunningAverageAreaSunspotsFunctionTime}
which shows first, that the auto-calibrated set of parameters we used throughout this article corresponds
to a very high structural similarity index $\textnormal{SSIM}\approx
0.82790361$ which is quite good and second, it shows that for widely different forecasts, some quite good (as measured by a high SSIM) and
some quite bad (low SSIM), we can have the same RMSE. We verified that these high/low SSIM values mean what they should
by examining visually the forecasted sunspot butterfly diagram and comparing it with the observed one. The conclusion is that
the RMSE metric is, as suspected, not a  perfect and unique way way to estimate the goodness of the forecasts.
\begin{figure}[htp]
\resizebox{\hsize}{!}{\includegraphics{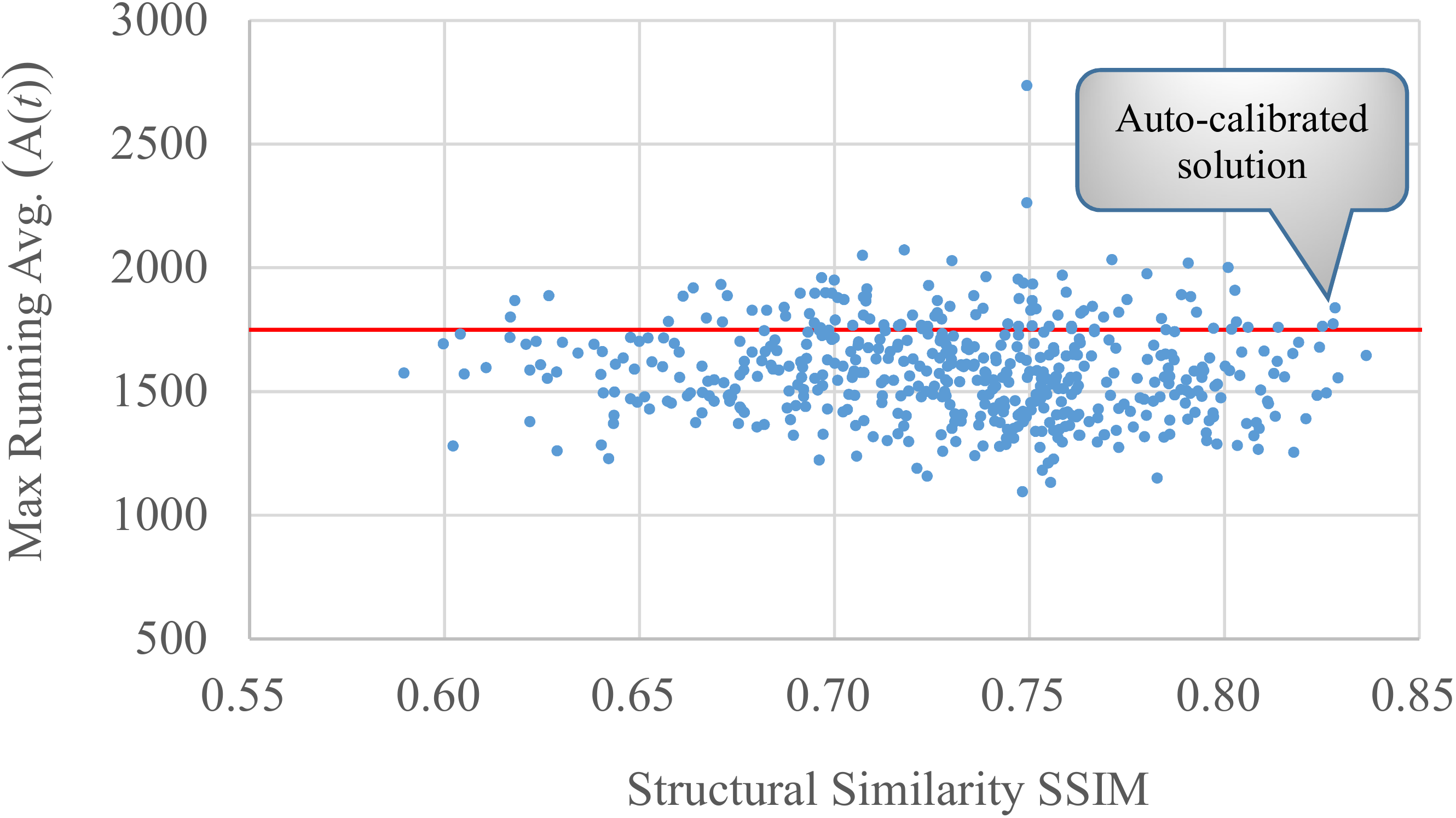}}
\caption{Maximum total sunspot area across the next forecasted cycle $\max(\overline{A(t)})$ against the structural similarity SSIM for random choices
of the forecasting algorithm's input parameters. This shows that similar quality forecasts,
at least in terms of structural similarity, can have quite different maximum sunspot areas over the next cycle. The red line marks the actual real maximum of the smoothed total sunspot area across the next cycle. This clearly shows that
markedly different forecasts of different quality (in space and time) can have a similar accurate maximum of the sunspot area.}
\label{StructuralSimilarity_versus_MaximumSunspotArea}
\end{figure}
The reason for the  difficulties when using the RMSE is that the butterfly diagram is not really a smooth two-dimensional surface, but a very irregular data set.

Finally, as most authors try to forecast the maximum sunspot area of the next or current cycle\footnote{In fact, most authors try to forecast the maximum
sunspot number or sunspot count, but the sunspot area is related quite closely to sunspot numbers.}
we also examined how  the forecast of the $\max(\overline{A(t)})$ over the next cycle (cycle 23) behaved against the SSIM
for random samples of the calibration input parameters. These results are depicted in Fig.\ \ref{StructuralSimilarity_versus_MaximumSunspotArea}.
Again, this shows that the calibration done using the first minimum of the average mutual information and the percentage of false
neighbours (the above-mentioned auto-calibration) seems to work, since this one shows a $\textnormal{SSIM}\approx
0.82790361$ and a forecasted total sunspot area  $\max(\overline{A(t)})$=\SI{1774.71},
which is quite good given that the observed maximum of the 24-point moving average total sunspot area for cycle 23 was \SI{1875.04}.
This seems to show that looking at a single number (the maximum solar activity over the next cycle) may not easily differentiate between
good and bad forecasts from the point of view of trying to match the full butterfly diagram.

% C:\Users\eurico\SunspotAnalysis\Papers\1502.07020v1.pdf
% here we can find the dates of minima to do what the referee asked:

%1). It is supposed to be a necessary step to evaluate the effectiveness of the method by the "prediction" of the past every cycle using the rest of the data. If the
%successful rate is high enough, it can be regarded as an effective prediction method to use. On contrast, the author directly used the data from 1874-1997 as the
%"training set" to predict the butterfly diagram since 1997 onwards.

%Actually, the sunspot emergence is cycle dependent. It is that the stronger cycles have higher latitude emergence of the sunspots. This is well investigated by Li
%et al. (2003, Solar Physics, 215, 99-109), Solanki et al. (2008, A&A, 483, 623), Jiang et al. (2011, A&A, 528, A82) and so on. It seems that their prediction
%results shown in Figures 6-9 only include the average information of sunspot emergence. Cycle 23 is a medium cycle. Hence, they can reproduce cycle 23 well, but
%fail in cycle 24, which is a weak cycle. They attribute the failure of cycle 24 to the time scope of the prediction. This is not justified in their pure
%mathematical prediction method, which is expected to predict enough long time series if the method is really effective.

\subsection{Effectiveness of the prediction per cycle}
\label{Effectiveness}

In  section \ref{originalforecast} we  attempted to predict cycle 23 and part of 24, using data up to cycle 22 inclusive.
However,  because the sunspot emergence and progression is cycle dependent \citep[see e.g.][]{1977BSolD1976...59V, 2001AJ....122.2115L,
Li2003, 2003ApJ...589..665H, 2008A&A...483..623S, 2011A&A...528A..82J, 2011SoPh..268..231I},
it is
quite fair to say that we may have just been lucky to reproduce cycle 23, given that it is a
medium cycle. After all, the method relies on training data, and if a cycle, e.g.\ cycle 24, is a weak cycle, then unless the training data
has visited that unusual or rarely visited part of the supposedly chaotic attractor, the forecast will inevitably fail.

To address this  question of the effectiveness of the method, we  ran an extra set of calculations. We  looped through all the cycles
and tried to forecast cycle $n$ using as the training set all cycles from the beginning (cycle 11) to $n-1$. To decide when one cycle finished and another
started, we  used the data in Tables 1 and 2 in \citet{2015LRSP...12....4H}, taking for the beginning of each cycle
the minimum in the 13-month running mean.
This analysis should be able to answer the question of the effectiveness of the
method with respect to the type  of the cycle (strong/medium/weak) we are trying to predict.

% \begin{table}[htb]
% \centering
% \caption{The structural similarity index SSIM, the RMSE\textsuperscript{*} of the
% moving average (24-point average) total sunspot area (sum over all latitudes) $\overline{A(t)}$ against the corresponding forecast.
% Given the values of the embedding parameters $J$ and $L$, we can only atempt forecasts from cycle 15 onwards, as the embedding vector
% ${\bf x}(s^n_m)$ would spill outside the training set for any forecast for cycle 14 or before. The higher the SSIM index, the more similar
% (in two dimensions) were the forecasts to the observed cycle.
% }
% \label{all_to_n_cycle_forecasts}
% \begin{tabular}{rrrrr}
% {\bf cycle number} & {\bf SSIM } & {\bf RMSE $\overline{A(t)}$}     & {\bf RMSE\textsuperscript{*} }  &  \\
% 15    &       0.7343234425    &        429.66         &        -      \\
% 16    &       0.7471987669    &        381.92         &        -      \\
% 17    &       0.7581530002    &        706.96         &        -      \\
% 18    &       0.7599822377    &        849.91         &        -      \\
% 19    &       0.6554588324    &        1113.51        &        -      \\
% 20    &       0.7964419219    &        172.46         &        -      \\
% 21    &       0.7419151079    &        633.14         &        -      \\
% 22    &       0.7772294107    &        566.30         &        -      \\
% 23    &       0.8293161160    &        138.87         &        -      \\
% 24    &       0.7257527617    &        623.67         &        -
% \end{tabular}
% \end{table}

\begin{figure}
\resizebox{\hsize}{!}{\includegraphics{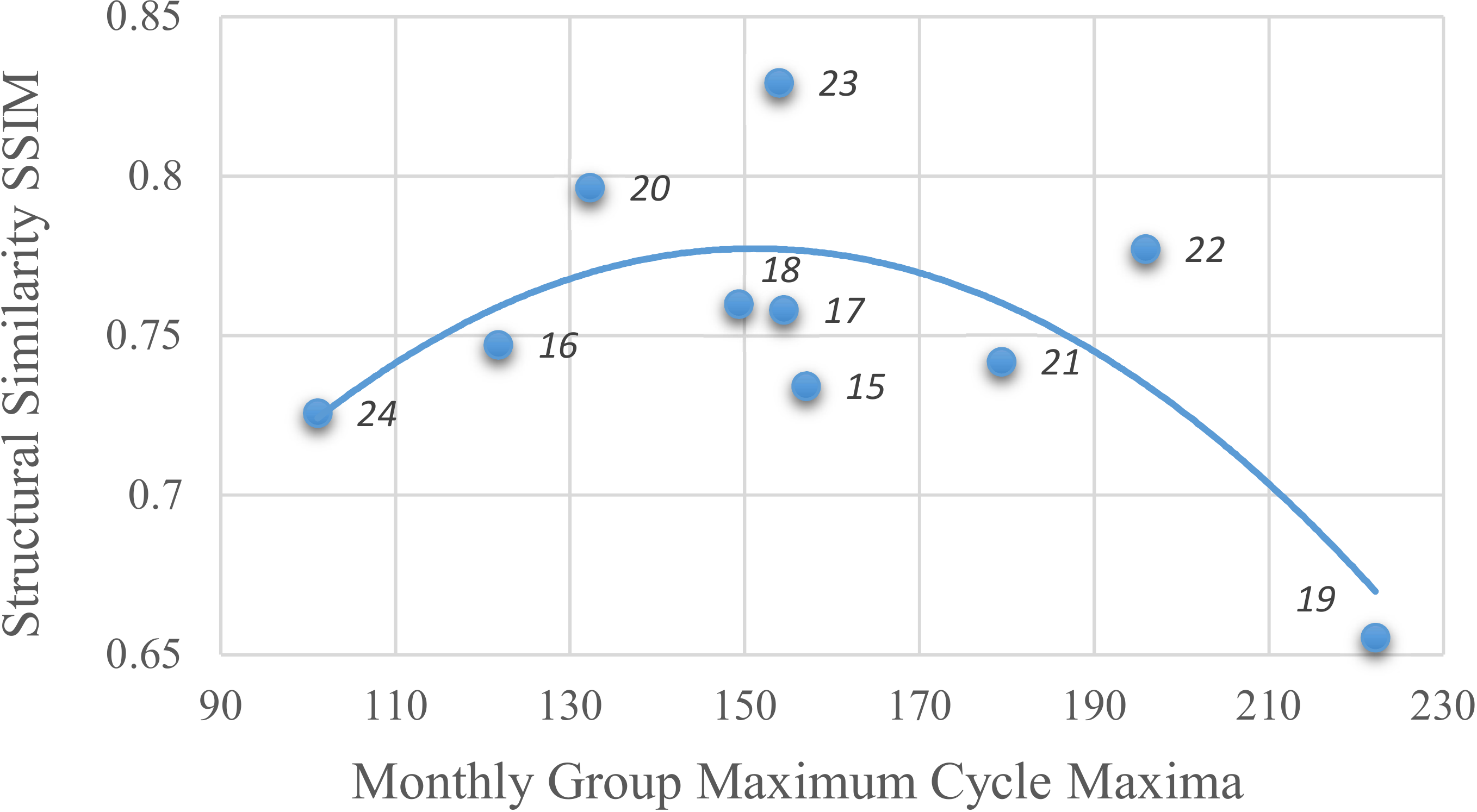}}
\caption{Structural similarity index SSIM of the forecast against the cycle strength, as given by the Monthly Group
Maximum sunspot number \citep{2015LRSP...12....4H}.  The results show that the method
is most effective for medium cycles.  We did not include attempts to forecast cycles 11 to 14 since the method requires   enough
past data, according to the auto-calibrated parameters $J$ and $L$; to forecast these cycles with this method we would need data prior to 1874.}
\label{SSIM_versus_cycle_maxima}
\end{figure}

The results of this analysis are depicted in Fig.\ \ref{SSIM_versus_cycle_maxima}, which seems to show two relationships.
First,  as the the number of data points increases, the predictive power
of the method as measured by the SSIM index seems to improve slightly. This is as expected, as the trajectory described by the ${\bf x}(s^n_m)$
embedding vectors will trace more and more the real chaotic attractor, as the time (or the size of the training set) increases.
Second,  the method described here seems to work better for medium cycles, as can be seen in the figure where we have superimposed a second-order fitting on the SSIM index against the strength of the cycle.
In particular, for the current cycle, cycle 24, which is a weak cycle, the forecast is not that good.

The method applied here works on the basis on finding the nearest neighbour
in the phase space of the embedding vectors ${\bf x}(s^n_m),$ and therefore it works better where there is a high density of neighbours. For weak and strong
cycles, the theoretical attractor will be poorly traced by the embedding vectors from real data. As these are the two most extreme cases, the
nearest neighbour method will be wrong more frequently. Unless more data is added, it is difficult to improve the performance of this method,
a clear limitation of just using sunspot data rather than all related physical data. Several
articles in the literature argue that one should use other data, e.g.\ geomagnetic precursors \citep{1982JGR....87.6153F, 1987GeoRL..14..632S,
1993SoPh..148..383T,
1999JGR...10422375H, 2006GeoRL..3318101H, 2007SoPh..243..205K, 2007ApJ...659..801C, 2009ApJ...694L..11W, 2016NatSR...621028N};
surface magnetic fields,  in particular polar fields
\citep{1978GeoRL...5..411S, 2005GeoRL..32.1104S, 2013PhRvL.111d1106M, 2014SSRv..186..325P}; polar faculae \citep{2012ApJ...753..146M, 2013ApJ...767L..25M};
and even helioseismological data \citep{2011Sci...333..993I, 2013ApJ...777..138I}. However,
to include more data for the training process is outside the scope of this article. Here we have limited ourselves to demonstrating
the possibility of forecasting using a pure statistical model, another approach to be added to the existing ones that use empirical relationships
\citep{2011A&A...528A..82J, 2016ApJ...823L..22C} or the magnetic field data \citep{0004-637X-792-1-12}.

However, we emphasise again the usefulness of attempting to do a forecast of the full butterfly diagram as opposed  to
just forecasting in one dimension (sunspot number or area full cycle) or zero dimensions (next sunspot maxima). The conclusion that
quite different spatial-temporal  forecasts can have very similar corresponding one or zero dimension results 
seems to be independent of the cycle to forecast, and if it is weak or strong.
%We show this in
%Fig.\ \ref{StructuralSimilarity_versus_MaximumSunspotArea_cycle24} where we calculate an analogous of Fig.\ \ref{StructuralSimilarity_versus_MaximumSunspotArea}
%for Monte Carlo simulations of forecasts of cycle 24. Although clearly we cannot forecast as well as we did for cycle 23, it shows that the
%conclusion that the same maxima can be accomplished with different similarities to the observed cycle, as measured by the SSIM index, is still valid.

A final question we may have is how far in the future can any method realistically predict. This depends fundamentally on whether the modulation
of the solar cycle, i.e. the variations in the period and amplitude of the sunspot area,
are a result of a chaotic attractor or simply a stochastic
variation on top of a basic cycle. This is not an easy question to answer, given the short time series available to us. 
The fact that our method, which assumes the existence of a chaotic attractor, can reproduce most of the qualitative features of the butterfly diagram, is
indicative that perhaps the solar cycle can be modelled by a chaotic attractor; however, this is just an indication, not  proof.
Some authors have tried to answer this question using different
numerical analysis \citep{1993A&A...274..497C, 1994A&A...290..983C,2010A&A...509A...5H,2012GeoRL..3910103L,2013A&A...550A...6H,2014RAA....14..104Z}, 
but given the small number  of data points it seems that at this stage we cannot be sure which hypothesis, 
a chaotic attractor or a stochastic variation on top of 
a periodic component, can be confirmed.

If we assume that the solar cycle can be modelled by low dimensional chaos, then to answer the 
 question of what is the horizon limit of any method, we can calculate the largest Lyapunov exponent on our system. If this is found to
be positive, then its inverse\footnote{The Lyapunov exponent of a dynamical system characterises the rate of separation of infinitesimally close trajectories
within the attractor \citep[see][and references therein]{1985PhyD...16..285W,1994PhLA..185...77K}. There are as many Lyapunov exponents as the number of dimensions of the system.
A first positive Lyapunov exponent
indicates chaos, i.e. infinitesimally close trajectories will separate at an exponential rate with time. The inverse of the
first positive Lyapunov exponent gives the limit or horizon of predictability in time - the Lyapunov time. By convention, the Lyapunov time
is defined as the time for the distance between infinitesimally close trajectories to increase by a factor of $e$.}
should give the upper limit of predictability. For our sunspot area data, we have calculated $\lambda_1 \approx  0.063$ bits month\textsuperscript{-1}, 
which implies a very short horizon of 15.9 months (or 1.33 years)\footnote{We used the \citet{1985PhyD...16..285W} method for the running average 
$\textnormal{RMSE}(\overline{A(t)})$ composed of 1865 points, using a time delay $\tau=42$ and an embedding dimension $D_e=4$, given by the
mutual information and false neighbourhood methods.}. There have been many previous attempts to
calculate the Lyapunov exponents \citep{
1990SoPh..127..405O, 
1991JGR....96.1705M,
1992AnGeo..10..759P,
1995AcApS..15...84Z,
1996A&A...310..646Z,
1997A&A...317..610C,
1998SoPh..178..423Z,
2006AGUFMSH21A0327C,
2008AnGeo..26..231B,
2009SoPh..255..301G,
2009A&A...506.1381C,
2009JGRA..114.1108C,
2012SunGe...7...75W,
2012PhyA..391.6287P, 
2013ApJ...779..108S,
2014RAA....14..104Z,
zhoufeng,
Zachilas2015}. Most values for the first or maximal Lyapunov exponent are around $0.01-0.03$ bits month\textsuperscript{-1},
which corresponds to 3 to 8 years. We note that these are mostly calculations of the Lyapunov exponent for sunspot number time series,
not the sunspot area spatial-temporal
series we use here, but the two are known to be related \citep{2006STIN...0620186W,2014RAA....14..104Z}. 
Given the dispersion of values, it seems that until longer time series are acquired, there is no guarantee that the Lyapunov exponent, and therefore
the horizon of predictability, can be calculated accurately.

% \begin{figure}[htp]
% \resizebox{\hsize}{!}{\includegraphics{StructuralSimilarity_versus_MaximumSunspotArea_cycle24-crop.pdf}}
% \caption{Maximum total sunspot area across the current cycle 24 $\max(\overline{A(t)})$ against the structural similarity SSIM for random choices
% of the forecasting algorithm's input parameters. The red line marks the actual real maximum of the smoothed total sunspot area across the next cycle.
% This clearly shows that
% markedly different forecast of different quality (in space and time) can have a similar maximum of the sunspot area.}
% \label{StructuralSimilarity_versus_MaximumSunspotArea_cycle24}
% \end{figure}

%check eurico's papers for lyapunov exponent calculation too!

% next paper do also the time of maximum and time of minimum

% do also https://www.mathworks.com/matlabcentral/fileexchange/43017-complex-wavelet-structural-similarity-index--cw-ssim-
% later in next paper

\section{Conclusion}
\label{conclusion}

 We attempt to predict the sunspot butterfly diagram in both space and time. We used a
prediction method based on the non-linear embeddings of spatial-temporal data series. This analysis
is in contrast with the usual
next cycle maximum sunspot number count,  the time of the next maximum/minimum, or the next cycle sunspot
number temporal shape, forecasts which are prevalent in the scientific literature.
This method has the advantage of being agnostic to the data source, i.e.\ it can be used in other settings, and of being  auto-calibrated, in
the sense that all the method's input parameters can be derived from the training set.

This analysis has been done
on publicly available sunspot area series from 1874 to 2015, which contain time and latitude information.
The analysis of the results shows that it is indeed possible to reproduce the overall `butterfly wing' shape and amplitude of the spatial-temporal
pattern of sunspots  using this method.
% It also seems to show
% that existing methods that just look at forecasting the (sunspot's count or area coverage) maximum for the next cycle or the time
% of the next maximum/minimum
% or attempt to predict the shape of the next cycle could
% be misleading.
However, to really know whether we are just reproducing an average cycle or if we have a method that truly can be called predictive, we also introduce the use of the so-called structural similarity index to compare the forecasted and observed cycles. The results of the comparison of this similarity index 
for all cycles show that indeed the method cannot predict weak or strong cycles, i.e. in its present form it does not have   predictive power.

In addition, our results show that different forecasts with distinct likeness to the observed future butterfly diagram
can have an  forecasted maximum and/or overall shape that is indistinguishable from the actual one. In summary, this type of forecast
opens up a new degree of freedom.
We believe that the structural similarity is therefore another tool to be used in addition to the root-mean-square
error and the correlation coefficient when evaluating the effectiveness of forecasts.

Finally, we believe that other methods, such as neural networks, should be explored to try to predict the full spatial-temporal butterfly diagram  (a forthcoming article on neural networks forecasts is in preparation). In addition  to sunspots, it would be interesting to explore
attempts to forecast the data series representing the
vertical magnetic flux through the solar photosphere as a function of  latitude/time, very high resolution data which goes back to 1975. In contrast to sunspot data, this data set  is smoother, so it should be a promising research area to explore. The data series representing bright spots in the Sun, called active region faculae, is also available and  should also be used 
to contrast with the sunspot analysis, as both sunspots and faculae are associated with magnetic fields and with driving the space weather.
Forecasting
the variation of the Sun's surface rotation rate with latitude and time from which a temporal average has been subtracted, known as torsional oscillations,
which shows a similar pattern to the sunspot butterfly diagram, could also be used as  input for these forecasting methods. Overall,
we believe that  forecasting in multiple dimensions, although harder computationally, should open many research paths within solar physics
and the area of space weather science.

% for magnetic field forecasts, check
% \citet{0004-637X-795-1-46}

% we must also try to do this magnetic field latitude/time one
%http://d29qn7q9z0j1p6.cloudfront.net/content/roypta/370/1970/3049/F1.large.jpg

% and faculae
% http://www.ngdc.noaa.gov/stp/solar/wfaculae.html

% and for helioseismological stuff
%http://soi.stanford.edu/press/GONG_MDI_03-00/presszone_col.gif

% and this one, is this real or simulated from MHD 3D code?
%https://www2.mps.mpg.de/projects/solar-mhd/pics/cameron/mer_vel_5001_.jpg

\begin{acknowledgements}
We thank Filipe C.\ Mena for very insightful discussions on the details of the non-linear embedding of data series in high dimensions
and Reza Tavakol for discussions on sunspot forecasting results. We  thank
Dr.\ David Hathaway for providing the data that this article is based upon. We also thank Nuno Peixinho for reviewing the draft article.
Finally, we thank an anonymous referee for comments that have helped improve this article.
\end{acknowledgements}

% for the bibliography, at the end
\bibliographystyle{aa} % style aa.bst
\bibliography{eurico} % your references Yourfile.bib

\begin{thebibliography}{169}
\expandafter\ifx\csname natexlab\endcsname\relax\def\natexlab#1{#1}\fi

\bibitem[{Abarbanel(1997)}]{abarbanel1997analysis}
Abarbanel, H. 1997, Analysis of Observed Chaotic Data, Institute for Nonlinear
  Science (Springer New York)

\bibitem[{{Abarbanel} {et~al.}(1993){Abarbanel}, {Brown}, {Sidorowich}, \&
  {Tsimring}}]{1993RvMP...65.1331A}
{Abarbanel}, H.~D.~I., {Brown}, R., {Sidorowich}, J.~J., \& {Tsimring}, L.~S.
  1993, Reviews of Modern Physics, 65, 1331

\bibitem[{{Abarbanel} \& {Gollub}(1996)}]{1996PhT....49k..86A}
{Abarbanel}, H.~D.~I. \& {Gollub}, J.~P. 1996, Physics Today, 49, 86

\bibitem[{{Aeyels}(1981)}]{zbMATH03744996}
{Aeyels}, D. 1981, {SIAM J. Control Optim.}, 19, 595

\bibitem[{{Arlt}(2008)}]{2008SoPh..247..399A}
{Arlt}, R. 2008, \solphys, 247, 399

\bibitem[{{Arlt}(2009)}]{2009SoPh..255..143A}
{Arlt}, R. 2009, \solphys, 255, 143

\bibitem[{{Arlt}(2011)}]{2011AN....332..805A}
{Arlt}, R. 2011, Astronomische Nachrichten, 332, 805

\bibitem[{{Arlt} \& {Abdolvand}(2011)}]{2011IAUS..273..286A}
{Arlt}, R. \& {Abdolvand}, A. 2011, in IAU Symposium, Vol. 273, Physics of Sun
  and Star Spots, ed. D.~{Prasad Choudhary} \& K.~G. {Strassmeier}, 286--289

\bibitem[{{Arlt} {et~al.}(2013){Arlt}, {Leussu}, {Giese}, {Mursula}, \&
  {Usoskin}}]{2013MNRAS.433.3165A}
{Arlt}, R., {Leussu}, R., {Giese}, N., {Mursula}, K., \& {Usoskin}, I.~G. 2013,
  \mnras, 433, 3165

\bibitem[{{Arlt} \& {Weiss}(2014)}]{2014SSRv..186..525A}
{Arlt}, R. \& {Weiss}, N. 2014, \ssr, 186, 525

\bibitem[{{Babayev}(2003)}]{2003A&AT...22..861B}
{Babayev}, E.~S. 2003, Astronomical and Astrophysical Transactions, 22, 861

\bibitem[{{Baranovski} {et~al.}(2008){Baranovski}, {Clette}, \&
  {Nollau}}]{2008AnGeo..26..231B}
{Baranovski}, A.~L., {Clette}, F., \& {Nollau}, V. 2008, Annales Geophysicae,
  26, 231

\bibitem[{{Beer} {et~al.}(1998){Beer}, {Tobias}, \&
  {Weiss}}]{1998SoPh..181..237B}
{Beer}, J., {Tobias}, S., \& {Weiss}, N. 1998, \solphys, 181, 237

\bibitem[{Bialonski {et~al.}(2015)Bialonski, Ansmann, \&
  Kantz}]{PhysRevE.92.042910}
Bialonski, S., Ansmann, G., \& Kantz, H. 2015, Phys. Rev. E, 92, 042910

\bibitem[{Bor{\v{s}}tnik~Bra{\v{c}}i{\v{c}}
  {et~al.}(2009)Bor{\v{s}}tnik~Bra{\v{c}}i{\v{c}}, Grabec, \&
  Govekar}]{BorstnikBracic2009}
Bor{\v{s}}tnik~Bra{\v{c}}i{\v{c}}, A., Grabec, I., \& Govekar, E. 2009, The
  European Physical Journal B, 69, 529

\bibitem[{{Brunet} {et~al.}(2012){Brunet}, {Vrscay}, \&
  {Wang}}]{2012ITIP...21.1488B}
{Brunet}, D., {Vrscay}, E.~R., \& {Wang}, Z. 2012, IEEE Transactions on Image
  Processing, 21, 1488

\bibitem[{{Cameron} \& {Sch{\"u}ssler}(2007)}]{2007ApJ...659..801C}
{Cameron}, R. \& {Sch{\"u}ssler}, M. 2007, \apj, 659, 801

\bibitem[{{Cameron} {et~al.}(2016){Cameron}, {Jiang}, \&
  {Sch{\"u}ssler}}]{2016ApJ...823L..22C}
{Cameron}, R.~H., {Jiang}, J., \& {Sch{\"u}ssler}, M. 2016, \apjl, 823, L22

\bibitem[{{Carbonell} {et~al.}(1993){Carbonell}, {Oliver}, \&
  {Ballester}}]{1993A&A...274..497C}
{Carbonell}, M., {Oliver}, R., \& {Ballester}, J.~L. 1993, \aap, 274, 497

\bibitem[{{Carbonell} {et~al.}(1994){Carbonell}, {Oliver}, \&
  {Ballester}}]{1994A&A...290..983C}
{Carbonell}, M., {Oliver}, R., \& {Ballester}, J.~L. 1994, \aap, 290, 983

\bibitem[{Cencini {et~al.}(2009)Cencini, Cecconi, \& Vulpiani}]{9814277657}
Cencini, M., Cecconi, F., \& Vulpiani, A. 2009, Chaos: From Simple Models to
  Complex Systems (Series on Advances in Statistical Mechanics) (World
  Scientific Publishing Company)

\bibitem[{{Charbonneau}(2001)}]{2001SoPh..199..385C}
{Charbonneau}, P. 2001, \solphys, 199, 385

\bibitem[{{Charbonneau}(2005)}]{2005SoPh..229..345C}
{Charbonneau}, P. 2005, \solphys, 229, 345

\bibitem[{{Charbonneau} {et~al.}(2004){Charbonneau}, {Blais-Laurier}, \&
  {St-Jean}}]{2004ApJ...616L.183C}
{Charbonneau}, P., {Blais-Laurier}, G., \& {St-Jean}, C. 2004, \apjl, 616, L183

\bibitem[{{Choi} {et~al.}(2011){Choi}, {Lee}, {Cho}, {Kwak}, {Cho}, {Park},
  {Kim}, {Baker}, {Reeves}, \& {Lee}}]{2011SpWea...9.6001C}
{Choi}, H.-S., {Lee}, J., {Cho}, K.-S., {et~al.} 2011, Space Weather, 9, 06001

\bibitem[{{Consolini} {et~al.}(2006){Consolini}, {Tozzi}, \& {de
  Michelis}}]{2006AGUFMSH21A0327C}
{Consolini}, G., {Tozzi}, R., \& {de Michelis}, P. 2006, AGU Fall Meeting
  Abstracts

\bibitem[{{Consolini} {et~al.}(2009){Consolini}, {Tozzi}, \& {de
  Michelis}}]{2009A&A...506.1381C}
{Consolini}, G., {Tozzi}, R., \& {de Michelis}, P. 2009, \aap, 506, 1381

\bibitem[{{Corn{\'e}lissen} {et~al.}(2009){Corn{\'e}lissen}, {Tarquini},
  {Perfetto}, {Otsuka}, {Gigolashvili}, \& {Halberg}}]{2009SunGe...4...55C}
{Corn{\'e}lissen}, G., {Tarquini}, R., {Perfetto}, F., {et~al.} 2009, Sun and
  Geosphere, 4, 55

\bibitem[{Covas \& Mena(2011)}]{Covas}
Covas, E. \& Mena, F. 2011, in Springer Proceedings in Mathematics, Vol.~1,
  Dynamics, Games and Science I, ed. M.~M. Peixoto, A.~A. Pinto, \& D.~A. Rand
  (Springer Berlin Heidelberg), 243--251

\bibitem[{{Covas} \& {Tavakol}(1997)}]{1997PhRvE..55.6641C}
{Covas}, E. \& {Tavakol}, R. 1997, \pre, 55, 6641

\bibitem[{{Covas} \& {Tavakol}(1999)}]{1999PhRvE..60.5435C}
{Covas}, E. \& {Tavakol}, R. 1999, \pre, 60, 5435

\bibitem[{{Covas} {et~al.}(2001){Covas}, {Tavakol}, {Ashwin}, {Tworkowski}, \&
  {Brooke}}]{2001Chaos..11..404C}
{Covas}, E., {Tavakol}, R., {Ashwin}, P., {Tworkowski}, A., \& {Brooke}, J.~M.
  2001, Chaos, 11, 404

\bibitem[{{Covas} {et~al.}(1997){Covas}, {Tworkowski}, {Brandenburg}, \&
  {Tavakol}}]{1997A&A...317..610C}
{Covas}, E., {Tworkowski}, A., {Brandenburg}, A., \& {Tavakol}, R. 1997, \aap,
  317, 610

\bibitem[{{Crosson} \& {Binder}(2009)}]{2009JGRA..114.1108C}
{Crosson}, I.~J. \& {Binder}, P.-M. 2009, Journal of Geophysical Research
  (Space Physics), 114, A01108

\bibitem[{{Darlington} {et~al.}(1995){Darlington}, {McGurk}, \&
  {Bray}}]{john1995chronicle}
{Darlington}, R., {McGurk}, P., \& {Bray}, J. 1995, The Chronicle of John of
  Worcester: The annals from 1067 to 1140 with the Gloucester interpolations
  and the continuation to 1141, Medieval Texts (Clarendon Press)

\bibitem[{{de Jager} \& {Usoskin}(2006)}]{2006JASTP..68.2053D}
{de Jager}, C. \& {Usoskin}, I. 2006, Journal of Atmospheric and
  Solar-Terrestrial Physics, 68, 2053

\bibitem[{{Deng} {et~al.}(2016){Deng}, {Li}, {Xiang}, \&
  {Dun}}]{2016AJ....151....2D}
{Deng}, L.~H., {Li}, B., {Xiang}, Y.~Y., \& {Dun}, G.~T. 2016, \aj, 151, 2

\bibitem[{{Diercke} {et~al.}(2015){Diercke}, {Arlt}, \&
  {Denker}}]{2015AN....336...53D}
{Diercke}, A., {Arlt}, R., \& {Denker}, C. 2015, Astronomische Nachrichten,
  336, 53

\bibitem[{{Dikpati} {et~al.}(2006){Dikpati}, {de Toma}, \&
  {Gilman}}]{2006GeoRL..33.5102D}
{Dikpati}, M., {de Toma}, G., \& {Gilman}, P.~A. 2006, \grl, 33, L05102

\bibitem[{{Duhau}(2003)}]{2003SoPh..213..203D}
{Duhau}, S. 2003, \solphys, 213, 203

\bibitem[{{Eddy}(1976)}]{1976Sci...192.1189E}
{Eddy}, J.~A. 1976, Science, 192, 1189

\bibitem[{{Eddy}(1983)}]{1983SoPh...89..195E}
{Eddy}, J.~A. 1983, \solphys, 89, 195

\bibitem[{EPSNRC(2012)}]{9780309265645}
EPSNRC. 2012, The Effects of Solar Variability on Earth's Climate: A Workshop
  Report, Committee on the Effects of Solar Variability on Earth's Climate and
  Space Studies Board and Division on Engineering and Physical Sciences and
  National Research Council (National Academies Press)

\bibitem[{{Ermolli} {et~al.}(2014){Ermolli}, {Shibasaki}, {Tlatov}, \& {van
  Driel-Gesztelyi}}]{2014SSRv..186..105E}
{Ermolli}, I., {Shibasaki}, K., {Tlatov}, A., \& {van Driel-Gesztelyi}, L.
  2014, \ssr, 186, 105

\bibitem[{{Feynman}(1982)}]{1982JGR....87.6153F}
{Feynman}, J. 1982, \jgr, 87, 6153

\bibitem[{Fraser \& Swinney(1986)}]{Fraser86}
Fraser, A.~M. \& Swinney, H.~L. 1986, Phys. Rev. A, 33, 1134

\bibitem[{{Friis-Christensen} \& {Lassen}(1991)}]{1991Sci...254..698F}
{Friis-Christensen}, E. \& {Lassen}, K. 1991, Science, 254, 698

\bibitem[{{Gleissberg} \& {Damboldt}(1971)}]{1971JBAA...81..270G}
{Gleissberg}, W. \& {Damboldt}, T. 1971, Journal of the British Astronomical
  Association, 81, 270

\bibitem[{{Greenkorn}(2009)}]{2009SoPh..255..301G}
{Greenkorn}, R.~A. 2009, \solphys, 255, 301

\bibitem[{{Hale}(1908)}]{1908ApJ....28..315H}
{Hale}, G.~E. 1908, \apj, 28, 315

\bibitem[{{Hanslmeier} \& {Braj{\v s}a}(2010)}]{2010A&A...509A...5H}
{Hanslmeier}, A. \& {Braj{\v s}a}, R. 2010, \aap, 509, A5

\bibitem[{{Hanslmeier} {et~al.}(2013){Hanslmeier}, {Braj{\v s}a}, {{\v
  C}alogovi{\'c}}, {Vr{\v s}nak}, {Ru{\v z}djak}, {Steinhilber}, {MacLeod},
  {Ivezi{\'c}}, \& {Skoki{\'c}}}]{2013A&A...550A...6H}
{Hanslmeier}, A., {Braj{\v s}a}, R., {{\v C}alogovi{\'c}}, J., {et~al.} 2013,
  \aap, 550, A6

\bibitem[{{Harvey}(1999)}]{1999ApJ...525C..60H}
{Harvey}, J. 1999, \apj, 525, 60

\bibitem[{{Hathaway}(2009)}]{2009SSRv..144..401H}
{Hathaway}, D.~H. 2009, \ssr, 144, 401

\bibitem[{{Hathaway}(2015{\natexlab{a}})}]{butterfly}
{Hathaway}, D.~H. 2015{\natexlab{a}}, {Sunspot Area Butterfly Diagram data}

\bibitem[{{Hathaway}(2015{\natexlab{b}})}]{2015LRSP...12....4H}
{Hathaway}, D.~H. 2015{\natexlab{b}}, Living Reviews in Solar Physics, 12
  [\eprint[arXiv]{1502.07020}]

\bibitem[{{Hathaway} {et~al.}(2003){Hathaway}, {Nandy}, {Wilson}, \&
  {Reichmann}}]{2003ApJ...589..665H}
{Hathaway}, D.~H., {Nandy}, D., {Wilson}, R.~M., \& {Reichmann}, E.~J. 2003,
  \apj, 589, 665

\bibitem[{{Hathaway} \& {Wilson}(2006)}]{2006GeoRL..3318101H}
{Hathaway}, D.~H. \& {Wilson}, R.~M. 2006, \grl, 33, L18101

\bibitem[{{Hathaway} {et~al.}(1999){Hathaway}, {Wilson}, \&
  {Reichmann}}]{1999JGR...10422375H}
{Hathaway}, D.~H., {Wilson}, R.~M., \& {Reichmann}, E.~J. 1999, \jgr, 104, 22

\bibitem[{{Ilonidis} {et~al.}(2013){Ilonidis}, {Zhao}, \&
  {Hartlep}}]{2013ApJ...777..138I}
{Ilonidis}, S., {Zhao}, J., \& {Hartlep}, T. 2013, \apj, 777, 138

\bibitem[{{Ilonidis} {et~al.}(2011){Ilonidis}, {Zhao}, \&
  {Kosovichev}}]{2011Sci...333..993I}
{Ilonidis}, S., {Zhao}, J., \& {Kosovichev}, A. 2011, Science, 333, 993

\bibitem[{{Ivanov} \& {Miletsky}(2011)}]{2011SoPh..268..231I}
{Ivanov}, V.~G. \& {Miletsky}, E.~V. 2011, \solphys, 268, 231

\bibitem[{{Jevti{\'c}} {et~al.}(2001){Jevti{\'c}}, {Schweitzer}, \&
  {Cellucci}}]{2001A&A...379..611J}
{Jevti{\'c}}, N., {Schweitzer}, J.~S., \& {Cellucci}, C.~J. 2001, \aap, 379,
  611

\bibitem[{Jiang \& Song(2011)}]{JiangS11}
Jiang, C. \& Song, F. 2011, JCP, 6, 1424

\bibitem[{{Jiang} {et~al.}(2011){Jiang}, {Cameron}, {Schmitt}, \&
  {Sch{\"u}ssler}}]{2011A&A...528A..82J}
{Jiang}, J., {Cameron}, R.~H., {Schmitt}, D., \& {Sch{\"u}ssler}, M. 2011,
  \aap, 528, A82

\bibitem[{{Jiang} {et~al.}(2007){Jiang}, {Chatterjee}, \&
  {Choudhuri}}]{2007MNRAS.381.1527J}
{Jiang}, J., {Chatterjee}, P., \& {Choudhuri}, A.~R. 2007, \mnras, 381, 1527

\bibitem[{{Kane}(1999)}]{1999SoPh..189..217K}
{Kane}, R.~P. 1999, \solphys, 189, 217

\bibitem[{{Kane}(2007)}]{2007SoPh..243..205K}
{Kane}, R.~P. 2007, \solphys, 243, 205

\bibitem[{{Kantz}(1994)}]{1994PhLA..185...77K}
{Kantz}, H. 1994, Physics Letters A, 185, 77

\bibitem[{Kantz \& Schreiber(1997)}]{opac-b1092652}
Kantz, H. \& Schreiber, T. 1997, Nonlinear time series analysis, Cambridge
  nonlinear science series (Cambridge, New York: Cambridge University Press),
  originally published: 1997

\bibitem[{{Kappenman}(2005)}]{2005SpWea...3.8C01K}
{Kappenman}, J.~G. 2005, Space Weather, 3, S08C01

\bibitem[{{Kennel} {et~al.}(1992){Kennel}, {Brown}, \&
  {Abarbanel}}]{1992PhRvA..45.3403K}
{Kennel}, M.~B., {Brown}, R., \& {Abarbanel}, H.~D.~I. 1992, \pra, 45, 3403

\bibitem[{{Kremliovsky}(1995)}]{1995SoPh..159..371K}
{Kremliovsky}, M.~N. 1995, \solphys, 159, 371

\bibitem[{{Kurths} \& {Ruzmaikin}(1990)}]{1990SoPh..126..407K}
{Kurths}, J. \& {Ruzmaikin}, A.~A. 1990, \solphys, 126, 407

\bibitem[{{Lassen} \& {Friis-Christensen}(1995)}]{1995JATP...57..835L}
{Lassen}, K. \& {Friis-Christensen}, E. 1995, Journal of Atmospheric and
  Terrestrial Physics, 57, 835

\bibitem[{{Layden} {et~al.}(1991){Layden}, {Fox}, {Howard}, {Sarajedini}, \&
  {Schatten}}]{1991SoPh..132....1L}
{Layden}, A.~C., {Fox}, P.~A., {Howard}, J.~M., {Sarajedini}, A., \&
  {Schatten}, K.~H. 1991, \solphys, 132, 1

\bibitem[{{Letellier} {et~al.}(2006){Letellier}, {Aguirre}, {Maquet}, \&
  {Gilmore}}]{2006A&A...449..379L}
{Letellier}, C., {Aguirre}, L.~A., {Maquet}, J., \& {Gilmore}, R. 2006, \aap,
  449, 379

\bibitem[{{Leussu} {et~al.}(2016){Leussu}, {Usoskin}, {Arlt}, \&
  {Mursula}}]{2016A&A...592A.160L}
{Leussu}, R., {Usoskin}, I.~G., {Arlt}, R., \& {Mursula}, K. 2016, \aap, 592,
  A160

\bibitem[{Li {et~al.}(2003)Li, Wang, Zhan, Yun, Liang, Zhao, \& Gu}]{Li2003}
Li, K., Wang, J., Zhan, L., {et~al.} 2003, Solar Physics, 215, 99

\bibitem[{{Li} {et~al.}(2001){Li}, {Yun}, \& {Gu}}]{2001AJ....122.2115L}
{Li}, K.~J., {Yun}, H.~S., \& {Gu}, X.~M. 2001, \aj, 122, 2115

\bibitem[{{Love} \& {Rigler}(2012)}]{2012GeoRL..3910103L}
{Love}, J.~J. \& {Rigler}, E.~J. 2012, \grl, 39, L10103

\bibitem[{{Ma{\~n}{\'e}}(1981)}]{1981LNM...898..230M}
{Ma{\~n}{\'e}}, R. 1981, Lecture Notes in Mathematics, Berlin Springer Verlag,
  898, 230

\bibitem[{Mandelj {et~al.}(2001)Mandelj, Grabec, \&
  Govekar}]{S0218127401003802}
Mandelj, S., Grabec, I., \& Govekar, E. 2001, International Journal of
  Bifurcation and Chaos, 11, 2731

\bibitem[{Mandelj {et~al.}(2004)Mandelj, Grabec, \&
  Govekar}]{S021812740401045X}
Mandelj, S., Grabec, I., \& Govekar, E. 2004, International Journal of
  Bifurcation and Chaos, 14, 2011

\bibitem[{{Martinerie} {et~al.}(1992){Martinerie}, {Albano}, {Mees}, \&
  {Rapp}}]{1992PhRvA..45.7058M}
{Martinerie}, J.~M., {Albano}, A.~M., {Mees}, A.~I., \& {Rapp}, P.~E. 1992,
  \pra, 45, 7058

\bibitem[{{Maunder}(1904)}]{1904MNRAS..64..747M}
{Maunder}, E.~W. 1904, \mnras, 64, 747

\bibitem[{{McIntosh} {et~al.}(2014){McIntosh}, {Wang}, {Leamon}, {Davey},
  {Howe}, {Krista}, {Malanushenko}, {Markel}, {Cirtain}, {Gurman}, {Pesnell},
  \& {Thompson}}]{2014ApJ...792...12M}
{McIntosh}, S.~W., {Wang}, X., {Leamon}, R.~J., {et~al.} 2014, \apj, 792, 12

\bibitem[{McIntosh {et~al.}(2014)McIntosh, Wang, Leamon, Davey, Howe, Krista,
  Malanushenko, Markel, Cirtain, Gurman, Pesnell, \&
  Thompson}]{0004-637X-792-1-12}
McIntosh, S.~W., Wang, X., Leamon, R.~J., {et~al.} 2014, The Astrophysical
  Journal, 792, 12

\bibitem[{{Mossman}(1989)}]{1989QJRAS..30...59M}
{Mossman}, J.~E. 1989, \qjras, 30, 59

\bibitem[{{Mu{\~n}oz-Jaramillo}
  {et~al.}(2013{\natexlab{a}}){Mu{\~n}oz-Jaramillo}, {Balmaceda}, \&
  {DeLuca}}]{2013PhRvL.111d1106M}
{Mu{\~n}oz-Jaramillo}, A., {Balmaceda}, L.~A., \& {DeLuca}, E.~E.
  2013{\natexlab{a}}, Physical Review Letters, 111, 041106

\bibitem[{{Mu{\~n}oz-Jaramillo}
  {et~al.}(2013{\natexlab{b}}){Mu{\~n}oz-Jaramillo}, {Dasi-Espuig},
  {Balmaceda}, \& {DeLuca}}]{2013ApJ...767L..25M}
{Mu{\~n}oz-Jaramillo}, A., {Dasi-Espuig}, M., {Balmaceda}, L.~A., \& {DeLuca},
  E.~E. 2013{\natexlab{b}}, \apjl, 767, L25

\bibitem[{{Mu{\~n}oz-Jaramillo} {et~al.}(2012){Mu{\~n}oz-Jaramillo}, {Sheeley},
  {Zhang}, \& {DeLuca}}]{2012ApJ...753..146M}
{Mu{\~n}oz-Jaramillo}, A., {Sheeley}, N.~R., {Zhang}, J., \& {DeLuca}, E.~E.
  2012, \apj, 753, 146

\bibitem[{{Mundt} {et~al.}(1991){Mundt}, {Maguire}, \&
  {Chase}}]{1991JGR....96.1705M}
{Mundt}, M.~D., {Maguire}, II, W.~B., \& {Chase}, R.~R.~P. 1991, \jgr, 96, 1705

\bibitem[{{Ng}(2016)}]{2016NatSR...621028N}
{Ng}, K.~K. 2016, Scientific Reports, 6, 21028

\bibitem[{{Ogurtsov}(2005{\natexlab{a}})}]{2005ARep...49..495O}
{Ogurtsov}, M.~G. 2005{\natexlab{a}}, Astronomy Reports, 49, 495

\bibitem[{{Ogurtsov}(2005{\natexlab{b}})}]{2005SoPh..231..167O}
{Ogurtsov}, M.~G. 2005{\natexlab{b}}, \solphys, 231, 167

\bibitem[{{Ossendrijver}(2003{\natexlab{a}})}]{2003A&ARv..11..287O}
{Ossendrijver}, M. 2003{\natexlab{a}}, \aapr, 11, 287

\bibitem[{{Ossendrijver}(2003{\natexlab{b}})}]{2003ASPC..286...97O}
{Ossendrijver}, M. 2003{\natexlab{b}}, in Astronomical Society of the Pacific
  Conference Series, Vol. 286, Current Theoretical Models and Future High
  Resolution Solar Observations: Preparing for ATST, ed. A.~A. {Pevtsov} \&
  H.~{Uitenbroek}, 97

\bibitem[{{Ossendrijver} \& {Covas}(2003)}]{2003IJBC....8.2327O}
{Ossendrijver}, M. \& {Covas}, E. 2003, International Journal of Bifurcation
  and Chaos, 8

\bibitem[{{Ostriakov} \& {Usoskin}(1990)}]{1990SoPh..127..405O}
{Ostriakov}, V.~M. \& {Usoskin}, I.~G. 1990, \solphys, 127, 405

\bibitem[{Packard {et~al.}(1980)Packard, Crutchfield, Farmer, \&
  Shaw}]{PhysRevLett.45.712}
Packard, N.~H., Crutchfield, J.~P., Farmer, J.~D., \& Shaw, R.~S. 1980, Phys.
  Rev. Lett., 45, 712

\bibitem[{Pan \& Billings(2008)}]{Pan2008}
Pan, Y. \& Billings, S. 2008, IEEE Transactions on Systems, Man, and
  Cybernetics, Part B (Cybernetics), 38, 846

\bibitem[{Parker(1979)}]{9780198512905}
Parker, E.~N. 1979, Cosmical Magnetic Fields: Their Origin and their Activity
  (The International Series of Monographs on Physics) (Oxford University Press)

\bibitem[{{Parlitz} \& {Merkwirth}(2000)}]{2000PhRvL..84.1890P}
{Parlitz}, U. \& {Merkwirth}, C. 2000, Physical Review Letters, 84, 1890

\bibitem[{{Pavlos} {et~al.}(1992){Pavlos}, {Dialetis}, {Kyriakou}, \&
  {Sarris}}]{1992AnGeo..10..759P}
{Pavlos}, G.~P., {Dialetis}, D., {Kyriakou}, G.~A., \& {Sarris}, E.~T. 1992,
  Annales Geophysicae, 10, 759

\bibitem[{{Pavlos} {et~al.}(2012){Pavlos}, {Karakatsanis}, \&
  {Xenakis}}]{2012PhyA..391.6287P}
{Pavlos}, G.~P., {Karakatsanis}, L.~P., \& {Xenakis}, M.~N. 2012, Physica A
  Statistical Mechanics and its Applications, 391, 6287

\bibitem[{Pesnell(2008)}]{pianoplot}
Pesnell, W.~D. 2008, Solar Physics, 252, 209

\bibitem[{{Pesnell}(2012)}]{2012SoPh..281..507P}
{Pesnell}, W.~D. 2012, \solphys, 281, 507

\bibitem[{{Petrie} {et~al.}(2014){Petrie}, {Petrovay}, \&
  {Schatten}}]{2014SSRv..186..325P}
{Petrie}, G.~J.~D., {Petrovay}, K., \& {Schatten}, K. 2014, \ssr, 186, 325

\bibitem[{{Pishkalo}(2014)}]{2014SoPh..289.1815P}
{Pishkalo}, M.~I. 2014, \solphys, 289, 1815

\bibitem[{{Polygiannakis} \& {Moussas}(1997)}]{1997jena.confE..55P}
{Polygiannakis}, J.~M. \& {Moussas}, X. 1997, in Joint European and National
  Astronomical Meeting, ed. J.~D. {Hadjidemetrioy} \& J.~H. {Seiradakis}, 55

\bibitem[{{Preminger} \& {Walton}(2007)}]{2007SoPh..240...17P}
{Preminger}, D.~G. \& {Walton}, S.~R. 2007, \solphys, 240, 17

\bibitem[{{Proctor}(2004)}]{2004A&G....45d..14P}
{Proctor}, M.~R.~E. 2004, Astronomy and Geophysics, 45, 4.14

\bibitem[{{Ruzmaikin}(1981)}]{1981ComAp...9...85R}
{Ruzmaikin}, A.~A. 1981, Comments on Astrophysics, 9, 85

\bibitem[{{Sabine}(1851)}]{1851RSPT..141..123S}
{Sabine}, E. 1851, Philosophical Transactions of the Royal Society of London
  Series I, 141, 123

\bibitem[{{Sabine}(1852)}]{1852RSPT..142..103S}
{Sabine}, E. 1852, Philosophical Transactions of the Royal Society of London
  Series I, 142, 103

\bibitem[{{Sauer} {et~al.}(1991){Sauer}, {Yorke}, \&
  {Casdagli}}]{1991JSP....65..579S}
{Sauer}, T., {Yorke}, J.~A., \& {Casdagli}, M. 1991, Journal of Statistical
  Physics, 65, 579

\bibitem[{{Schatten}(2005)}]{2005GeoRL..3221106S}
{Schatten}, K. 2005, \grl, 32, L21106

\bibitem[{{Schatten} {et~al.}(1996){Schatten}, {Myers}, \&
  {Sofia}}]{1996GeoRL..23..605S}
{Schatten}, K., {Myers}, D.~J., \& {Sofia}, S. 1996, \grl, 23, 605

\bibitem[{{Schatten} {et~al.}(1978){Schatten}, {Scherrer}, {Svalgaard}, \&
  {Wilcox}}]{1978GeoRL...5..411S}
{Schatten}, K.~H., {Scherrer}, P.~H., {Svalgaard}, L., \& {Wilcox}, J.~M. 1978,
  \grl, 5, 411

\bibitem[{{Schatten} \& {Sofia}(1987)}]{1987GeoRL..14..632S}
{Schatten}, K.~H. \& {Sofia}, S. 1987, \grl, 14, 632

\bibitem[{{Schreiber}(1999)}]{1999PhR...308....1S}
{Schreiber}, T. 1999, \physrep, 308, 1

\bibitem[{{Schwabe}(1844)}]{1844AN.....21..233S}
{Schwabe}, M. 1844, Astronomische Nachrichten, 21, 233

\bibitem[{{Sello}(2001)}]{2001A&A...377..312S}
{Sello}, S. 2001, \aap, 377, 312

\bibitem[{{Senthamizh Pavai} {et~al.}(2015){Senthamizh Pavai}, {Arlt},
  {Dasi-Espuig}, {Krivova}, \& {Solanki}}]{2015A&A...584A..73S}
{Senthamizh Pavai}, V., {Arlt}, R., {Dasi-Espuig}, M., {Krivova}, N.~A., \&
  {Solanki}, S.~K. 2015, \aap, 584, A73

\bibitem[{{Shapoval} {et~al.}(2013){Shapoval}, {Le Mou{\"e}l}, {Courtillot}, \&
  {Shnirman}}]{2013ApJ...779..108S}
{Shapoval}, A., {Le Mou{\"e}l}, J.~L., {Courtillot}, V., \& {Shnirman}, M.
  2013, \apj, 779, 108

\bibitem[{{Singh} {et~al.}(2011){Singh}, {Siingh}, \&
  {Singh}}]{2011AtmEn..45.3806S}
{Singh}, A.~K., {Siingh}, D., \& {Singh}, R.~P. 2011, Atmospheric Environment,
  45, 3806

\bibitem[{{Solanki} {et~al.}(2008){Solanki}, {Wenzler}, \&
  {Schmitt}}]{2008A&A...483..623S}
{Solanki}, S.~K., {Wenzler}, T., \& {Schmitt}, D. 2008, \aap, 483, 623

\bibitem[{{Spiegel}(2009)}]{2009SSRv..144...25S}
{Spiegel}, E.~A. 2009, \ssr, 144, 25

\bibitem[{{Steinhilber} {et~al.}(2012){Steinhilber}, {Abreu}, {Beer},
  {Brunner}, {Christl}, {Fischer}, {Heikkila}, {Kubik}, {Mann}, {McCracken},
  {Miller}, {Miyahara}, {Oerter}, \& {Wilhelms}}]{2012PNAS..109.5967S}
{Steinhilber}, F., {Abreu}, J.~A., {Beer}, J., {et~al.} 2012, Proceedings of
  the National Academy of Science, 109, 5967

\bibitem[{{Stephenson} \& {Arny}(1980)}]{1980AmJPh..48..258S}
{Stephenson}, F.~R. \& {Arny}, T.~T. 1980, American Journal of Physics, 48, 258

\bibitem[{{Stuiver}(1980)}]{1980Natur.286..868S}
{Stuiver}, M. 1980, \nat, 286, 868

\bibitem[{{Stuiver} \& {Quay}(1980)}]{1980Sci...207...11S}
{Stuiver}, M. \& {Quay}, P.~D. 1980, Science, 207, 11

\bibitem[{{Suess}(1979)}]{1979P&SS...27.1001S}
{Suess}, S.~T. 1979, \planss, 27, 1001

\bibitem[{{Svalgaard} {et~al.}(2005){Svalgaard}, {Cliver}, \&
  {Kamide}}]{2005GeoRL..32.1104S}
{Svalgaard}, L., {Cliver}, E.~W., \& {Kamide}, Y. 2005, \grl, 32, L01104

\bibitem[{{Takens}(1981)}]{1981LNM...898..366T}
{Takens}, F. 1981, Lecture Notes in Mathematics, Berlin Springer Verlag, 898,
  366

\bibitem[{{Tavakol} \& {Covas}(1999)}]{1999ASPC..178..173T}
{Tavakol}, R. \& {Covas}, E. 1999, in Astronomical Society of the Pacific
  Conference Series, Vol. 178, Stellar Dynamos: Nonlinearity and Chaotic Flows,
  ed. M.~{Nunez} \& A.~{Ferriz-Mas}, 173

\bibitem[{{Tavakol}(1978)}]{1978Natur.276..802T}
{Tavakol}, R.~K. 1978, \nat, 276, 802

\bibitem[{{Thompson}(1993)}]{1993SoPh..148..383T}
{Thompson}, R.~J. 1993, \solphys, 148, 383

\bibitem[{{Turner}(2006)}]{2006GMS...165..367T}
{Turner}, R.~E. 2006, Washington DC American Geophysical Union Geophysical
  Monograph Series, 165, 367

\bibitem[{{Usoskin} \& {Mursula}(2003)}]{2003SoPh..218..319U}
{Usoskin}, I.~G. \& {Mursula}, K. 2003, \solphys, 218, 319

\bibitem[{{Usoskin} {et~al.}(2009){Usoskin}, {Mursula}, {Arlt}, \&
  {Kovaltsov}}]{2009ApJ...700L.154U}
{Usoskin}, I.~G., {Mursula}, K., {Arlt}, R., \& {Kovaltsov}, G.~A. 2009, \apjl,
  700, L154

\bibitem[{{Vitinskij}(1977)}]{1977BSolD1976...59V}
{Vitinskij}, Y.~I. 1977, Byulletin Solnechnye Dannye Akademie Nauk SSSR, 1976,
  59

\bibitem[{{Waldmeier}(1961)}]{1961says.book.....W}
{Waldmeier}, M. 1961, {The sunspot-activity in the years 1610-1960}
  (Schulthess)

\bibitem[{{Wang} \& {Sheeley}(2009)}]{2009ApJ...694L..11W}
{Wang}, Y.-M. \& {Sheeley}, N.~R. 2009, \apjl, 694, L11

\bibitem[{Wang {et~al.}(2004)Wang, Bovik, Sheikh, \&
  Simoncelli}]{Wang04imagequality}
Wang, Z., Bovik, A.~C., Sheikh, H.~R., \& Simoncelli, E.~P. 2004, IEEE
  TRANSACTIONS ON IMAGE PROCESSING, 13, 600

\bibitem[{{Weiss}(1988)}]{1988ssgv.conf...69W}
{Weiss}, N.~O. 1988, in Secular Solar and Geomagnetic Variations in the Last
  10,000 Years,, ed. F.~R. {Stephenson} \& A.~W. {Wolfendale}, 69--78

\bibitem[{{Weiss}(1990)}]{1990RSPTA.330..617W}
{Weiss}, N.~O. 1990, Philosophical Transactions of the Royal Society of London
  Series A, 330, 617

\bibitem[{{Weiss} \& {Tobias}(2016)}]{2016MNRAS.456.2654W}
{Weiss}, N.~O. \& {Tobias}, S.~M. 2016, \mnras, 456, 2654

\bibitem[{{Werner}(2012)}]{2012SunGe...7...75W}
{Werner}, R. 2012, Sun and Geosphere, 7, 75

\bibitem[{{West} {et~al.}(2013){West}, {Seaton}, {Dominique}, {Berghmans},
  {Nicula}, {Pylyser}, {Stegen}, \& {De Keyser}}]{2013EGUGA..1510865W}
{West}, M., {Seaton}, D., {Dominique}, M., {et~al.} 2013, in EGU General
  Assembly Conference Abstracts, Vol.~15, EGU General Assembly Conference
  Abstracts, EGU2013--10865

\bibitem[{Whitney(1936)}]{key1503303m}
Whitney, H. 1936, Ann. Math. (2), 37, 645, mR:1503303. Zbl:0015.32001.
  JFM:62.1454.01.

\bibitem[{{Wilkinson} {et~al.}(2000){Wilkinson}, {Shea}, \&
  {Smart}}]{2000AdSpR..26...27W}
{Wilkinson}, D.~C., {Shea}, M.~A., \& {Smart}, D.~F. 2000, Advances in Space
  Research, 26, 27

\bibitem[{{Wilson}(2006)}]{Wilson2006}
{Wilson}, I.~R.~G. 2006, in Secular Solar and Geomagnetic Variations in the
  Last 10,000 Years,

\bibitem[{{Wilson} {et~al.}(2008){Wilson}, {Carter}, \&
  {Waite}}]{2008PASA...25...85W}
{Wilson}, I.~R.~G., {Carter}, B.~D., \& {Waite}, I.~A. 2008, \pasa, 25, 85

\bibitem[{{Wilson}(1994)}]{1994ssac.book.....W}
{Wilson}, P.~R. 1994, {Solar and stellar activity cycles} (Cambridge University
  Press)

\bibitem[{{Wilson}(1987)}]{1987JGR....9210101W}
{Wilson}, R.~M. 1987, \jgr, 92, 10

\bibitem[{{Wilson} \& {Hathaway}(2006)}]{2006STIN...0620186W}
{Wilson}, R.~M. \& {Hathaway}, D.~H. 2006, NASA STI/Recon Technical Report N, 6

\bibitem[{{Wittmann}(1978)}]{1978A&A....66...93W}
{Wittmann}, A. 1978, \aap, 66, 93

\bibitem[{{Wolf} {et~al.}(1985){Wolf}, {Swift}, {Swinney}, \&
  {Vastano}}]{1985PhyD...16..285W}
{Wolf}, A., {Swift}, J.~B., {Swinney}, H.~L., \& {Vastano}, J.~A. 1985, Physica
  D Nonlinear Phenomena, 16, 285

\bibitem[{{Wolf}(1852)}]{1852MNRAS..13...29W}
{Wolf}, M. 1852, \mnras, 13, 29

\bibitem[{{Wolf}(1859)}]{1859MNRAS..19...85W}
{Wolf}, R. 1859, \mnras, 19, 85

\bibitem[{{Yallop} \& {Hohenkerk}(1980)}]{1980SoPh...68..303Y}
{Yallop}, B.~D. \& {Hohenkerk}, C.~Y. 1980, \solphys, 68, 303

\bibitem[{Zachilas \& Gkana(2015)}]{Zachilas2015}
Zachilas, L. \& Gkana, A. 2015, Solar Physics, 290, 1457

\bibitem[{{Zhang}(1995)}]{1995AcApS..15...84Z}
{Zhang}, Q. 1995, Acta Astrophysica Sinica, 15, 84

\bibitem[{{Zhang}(1996)}]{1996A&A...310..646Z}
{Zhang}, Q. 1996, \aap, 310, 646

\bibitem[{{Zhang}(1998)}]{1998SoPh..178..423Z}
{Zhang}, Q. 1998, \solphys, 178, 423

\bibitem[{Zhou \& Feng(2014)}]{zhoufeng}
Zhou, S. \& Feng, Y. 2014, International Journal of u- and e- Service, Science
  and Technology, 7, 73

\bibitem[{{Zhou} {et~al.}(2014){Zhou}, {Feng}, {Wu}, {Li}, \&
  {Liu}}]{2014RAA....14..104Z}
{Zhou}, S., {Feng}, Y., {Wu}, W.-Y., {Li}, Y., \& {Liu}, J. 2014, Research in
  Astronomy and Astrophysics, 14, 104

\end{thebibliography}

\end{document}